\documentclass{article}
\usepackage[dvipsnames]{xcolor}  
\usepackage{pbg_template_purple}
\usepackage{pifont}
\usepackage{parskip}

\usepackage[utf8]{inputenc}
\usepackage[T1]{fontenc}
\definecolor{MyPurple}{HTML}{6B67EE}
\definecolor{MyPink}{HTML}{ca5670}

\providecolor{PennBlue}{HTML}{01256E}

\makeatletter
\renewcommand\paragraph{\@startsection
  {paragraph}{4}{\z@}
  {\parskip}
  {-1em}
  {\normalfont\normalsize\bfseries\color{PennBlue}}}

\@ifundefined{subsubsection}{%
  \newcommand\subsubsection{\@startsection
    {subsubsection}{3}{\z@}
    {-2.5ex\@plus -1ex \@minus -.25ex}%
    {1.25ex \@plus .25ex}%
    {\normalfont\normalsize\bfseries\color{PennBlue}}}%
}{%
  \renewcommand\subsubsection{\@startsection
    {subsubsection}{3}{\z@}
    {-2.5ex\@plus -1ex \@minus -.25ex}%
    {1.25ex \@plus .25ex}%
    {\normalfont\normalsize\bfseries\color{PennBlue}}}%
}
\makeatother

\usepackage[colorlinks=true, linkcolor=MyPurple, citecolor=MyPurple, urlcolor=MyPurple]{hyperref}
\usepackage{natbib}
\usepackage{booktabs}       
\usepackage{amsfonts}       
\usepackage{nicefrac}       
\usepackage{microtype}      
\usepackage{amsmath}
\usepackage{amssymb}
\usepackage{graphicx}
\usepackage{tikz}
\usepackage{subcaption}
\usepackage{colortbl}
\usepackage{algorithm}
\usepackage{subcaption}
\usepackage{tabularx}
\usepackage{amsmath,amssymb,amsthm,bbm}
\usepackage{multirow}
\usepackage{adjustbox}
\usepackage{algorithm}
\usepackage{algpseudocode}
\usepackage{amsmath}
\usepackage[noEnd=false]{algpseudocodex}
\usepackage[toc,page]{appendix} 
\usepackage{tocloft} 
\usepackage{titlesec}
\usepackage{booktabs}
\usepackage{threeparttable}
\usepackage{enumitem}
\setlist[enumerate]{left=10pt}
\usepackage[font=small]{caption}
\usepackage{wrapfig}
\usepackage{listings}
\usepackage{booktabs}
\usepackage{mdframed}
\usepackage{thm-restate}

\usepackage{booktabs}
\usepackage{multirow}
\usepackage{array}
\usepackage{thmtools}

\title{mRNAutilus: Multi-Objective-Guided Discrete Generation of mRNA with Optimized Therapeutic Properties}

\author{Sawan Patel,\textsuperscript{1,*}
    Sophia Tang,\textsuperscript{2,*}
    Yesol Kim,\textsuperscript{3,*}
    Yinuo Zhang,\textsuperscript{3,4} 
    Divya Srijay,\textsuperscript{3}\hspace{1cm}
    Ping-Jung Lin,\textsuperscript{5}
    Shambhavi Shubham,\textsuperscript{5}
    Fengmei Pi,\textsuperscript{5}
    Cedric Wu,\textsuperscript{5}
    Sherwood Yao\textsuperscript{1,\dag}
    Pranam Chatterjee\textsuperscript{2,3,\dag}
    
    \vspace{1em}
    \normalfont \small
    \textsuperscript{1}Atom Bioworks, Inc.\\
    \textsuperscript{2}Department of Computer and Information Science, University of Pennsylvania\\
    \textsuperscript{3}Department of Bioengineering, University of Pennsylvania\\
    \textsuperscript{4}Center of Computational Biology, Duke-NUS Medical School\\ 
    \textsuperscript{5}GenScript USA, Inc.\\
    
    \vspace{0.5em}
    \textit{\textsuperscript{*}Equal contribution}
    
    \vspace{0.5em}
    \textbf{Correspondence:} \href{mailto:s.yao@atombioworks.com}{\texttt{s.yao@atombioworks.com}} and \href{mailto:pranam@seas.upenn.edu}{\texttt{pranam@seas.upenn.edu}} \\
    \textbf{Interface:} \href{https://autona.atombio.ai}{\texttt{https://autona.atombio.ai}}
}
\begin{document}

\maketitle

\begin{abstract}
    Therapeutic mRNA design requires coordinating multiple interacting sequence features across the full transcript, where codon usage, untranslated regions (UTRs), and their coupling jointly determine stability, translation efficiency, and protein expression. Here, we present \textbf{mRNA} generation via \textbf{u}nrolled \textbf{t}rajectories and \textbf{i}nformed \textbf{l}atent \textbf{u}pdate\textbf{s} (\textbf{mRNAutilus}), a framework for simultaneous codon optimization and \textit{de novo} UTR design directly from sequence. mRNAutilus combines a masked discrete diffusion model trained on millions of full-length mRNAs with Monte Carlo Tree Guidance to generate Pareto-efficient sequences under multiple functional objectives, using lightweight regressors over model embeddings to predict half-life, translation efficiency, and protein abundance. Unlike recent methods that design coding sequences and UTRs separately or rely on post hoc assembly and screening, mRNAutilus generates complete transcripts in a single process optimized across properties. Across diverse targets, zero-shot mRNAs encoding \textit{P. pyralis} luciferase achieve over 400-fold higher expression than wild-type and outperform commercial and machine learning-designed baselines, including zero-shot generative approaches. Zero-shot SARS-CoV-2 Spike mRNAs exceed clinically used and commercial constructs and match or surpass lab-optimized designs with improved durability. We further demonstrate generality in therapeutic settings, including prime editing (PEMax) and programmable proteome modulation, where mRNAutilus-designed constructs enhance expression of peptide-guided E3 ligases (uAbs) for $\beta$-catenin degradation. These results establish a sequence-based, multi-objective framework for generating functional mRNAs tailored to diverse biological applications.
\end{abstract}

\section{Introduction}
Messenger RNA (mRNA) has emerged as a transformative therapeutic modality, with applications spanning vaccines, protein replacement therapies, gene editing, and diagnostics \citep{pardi2018mrna, pardi2020recent, jackson2020promise, zhang2019advances, barbier2022clinical, rohner2022unlocking, popovitz2023gene}. The COVID-19 pandemic underscored this potential, as novel mRNA vaccines were developed, manufactured, and distributed globally within months \citep{park2021mrna}, and several mRNA candidates have since advanced into Phase 3 clinical trials \citep{soens2025phase, weber2024individualised, wilson2023efficacy}. Yet mRNAs are inherently unstable and susceptible to nuclease-mediated degradation \citep{ross1995mrna}, thereby limiting protein expression and therapeutic efficacy. Optimizing mRNA for stability, translation, and expression is therefore a central design objective, and one complicated by the fact that these properties are jointly determined by codon choice in the coding sequence (CDS) \citep{mauro2014critical}, the 5' and 3' untranslated regions (UTRs) that govern translation initiation and stability \citep{xiao2020facilitating, leppek2018functional, mauger2019mrna, ng20175}, and the inter-region interactions coupling them \citep{lin2015rna, lin2022deciphering, leppek2022combinatorial}.

\begin{figure*}
    \centering
    \includegraphics[width=\linewidth]{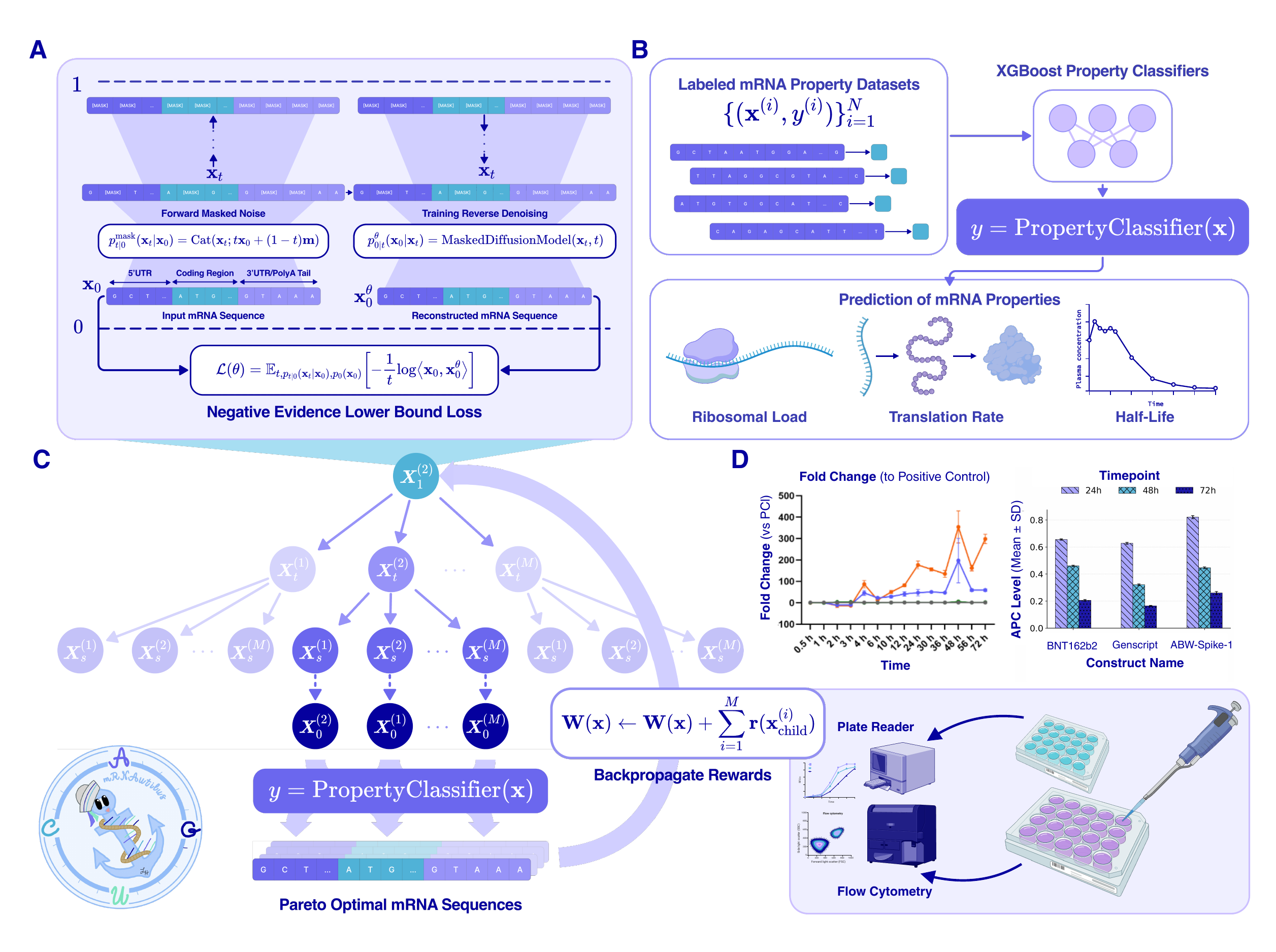}
    \vspace{-10pt}
    \caption{\textbf{Overview of mRNAutilus.} \textbf{(A)} Masked diffusion model pretraining and \textit{de novo} mRNA sequence generation. \textbf{(B)} mRNA function and property prediction using XGBoost regression analysis from model embeddings. \textbf{(C)} Multi-objective codon optimization and UTR generation using Monte Carlo Tree Guidance, which leverages tree search to optimize a set of Pareto-optimal sequences. \textbf{(D)} \textit{In vitro} experiments demonstrate that zero-shot mRNAs generated with mRNAutilus achieve higher expression and half-life than those produced by other machine-learning-driven methods and commercial mRNAs.}
    \label{fig:overview}
\end{figure*}

Existing mRNA design strategies fall short of this multi-property challenge. Traditional codon-optimization approaches maximize the codon adaptation index via species-specific codon-usage tables \citep{sharp1987codon}, but treat codons in isolation and ignore broader sequence context \citep{mauro2014critical, fu2020codon, leppek2022combinatorial, rosa2022maximizing, gebre2022optimization, kim2022modifications}. \textit{In-silico} screening with predictive models \citep{li2025mrna, zheng2025predicting} and graph-based or language-model approaches \citep{zhang2023algorithm, yazdani2025helm, zhang2025mrna2vec, wood2025helix, patel2025evoflow} similarly offer useful fitness estimates but do not directly generate optimized sequences. Recent generative efforts have targeted individual components: stability-guided 3'UTRs \citep{morrowmdd}, 5'UTRs \citep{chu20245, castillo2024optimizing}, or component-wise CDS and UTR models \citep{zhang2025deep}. However, piecewise design misses the cross-component couplings that collectively shape mRNA fitness \citep{lin2015rna, lin2022deciphering, metkar2024tailor}, and stability alone does not determine efficacy. Reward-guided autoregressive methods have recently been applied for this problem but demonstrate inconclusive results \textit{in silico} \citep{li2026mrna}. Ensemble-based workflows integrating independent modules for modeling mRNAs have similarly been explored, yet underperform on zero-shot design tasks \citep{kim2026vaxlab}.

Addressing this gap requires generative models capable of simultaneously designing the full transcript under multiple objectives. Masked discrete diffusion models (MDMs) provide a natural architecture: they support flexible, any-order generation, capture bidirectional dependencies \citep{shi2024simplified, austin2021structured, Sahoo2024}, and have proven effective for biomolecular design in proteins and peptides. On their own, however, MDMs sample from the data distribution and must be steered toward desired properties. Multi-objective guidance methods have enabled joint optimization of competing therapeutic properties with minimal tradeoffs \citep{abeer2024multi, jin2020multi}, and Monte Carlo Tree Guidance (MCTG) \citep{tang2025peptune, tang2025tr2} in particular couples a pre-trained discrete diffusion backbone with arbitrary reward functions, having been shown to navigate the vast design space of non-canonical and cyclic peptides toward Pareto-optimal sequences.

In this work, we present \textbf{mRNA} generation via \textbf{u}nrolled \textbf{t}rajectories and \textbf{i}nformed \textbf{l}atent \textbf{u}pdate\textbf{s} (\textbf{mRNAutilus}) (Figure~\ref{fig:overview}), a multi-objective-guided MDM that jointly performs codon optimization and full UTR design, steered by lightweight regressors over model embeddings predicting half-life, translation efficiency, and protein abundance. A hybrid tokenization scheme supports UTR-only, CDS-only, or whole-transcript design. Across diverse targets, zero-shot mRNAutilus designs for \textit{P. pyralis} luciferase express over 400-fold above wild-type and up to 6-fold above other zero-shot machine-learning-designed mRNAs across cancer cell lines; zero-shot SARS-CoV-2 Spike express nearly 2-fold above commercially-optimized constructs and achieve more durable expression than lab-in-the-loop designs, while designed PEMax mRNA sequences demonstrate improved editing efficiency over commercially-optimized designs. To demonstrate generality beyond reporters and antigens, we further apply mRNAutilus to peptide-guided E3 ubiquitin ligases, termed ubiquibodies (uAbs) \citep{portnoff2014ubiquibodies, brixi2023salt, bhat2025novo, chen2025target}, where optimized constructs drive enhanced intracellular expression and proteasome-dependent degradation of endogenous $\beta$-catenin. Together, these results establish mRNAutilus as a unified, sequence-based framework for programmable mRNA optimization spanning vaccines, reporters, and intracellular degraders.

\section{Results}

\begin{figure}[t]
    \centering
    \includegraphics[width=\linewidth]{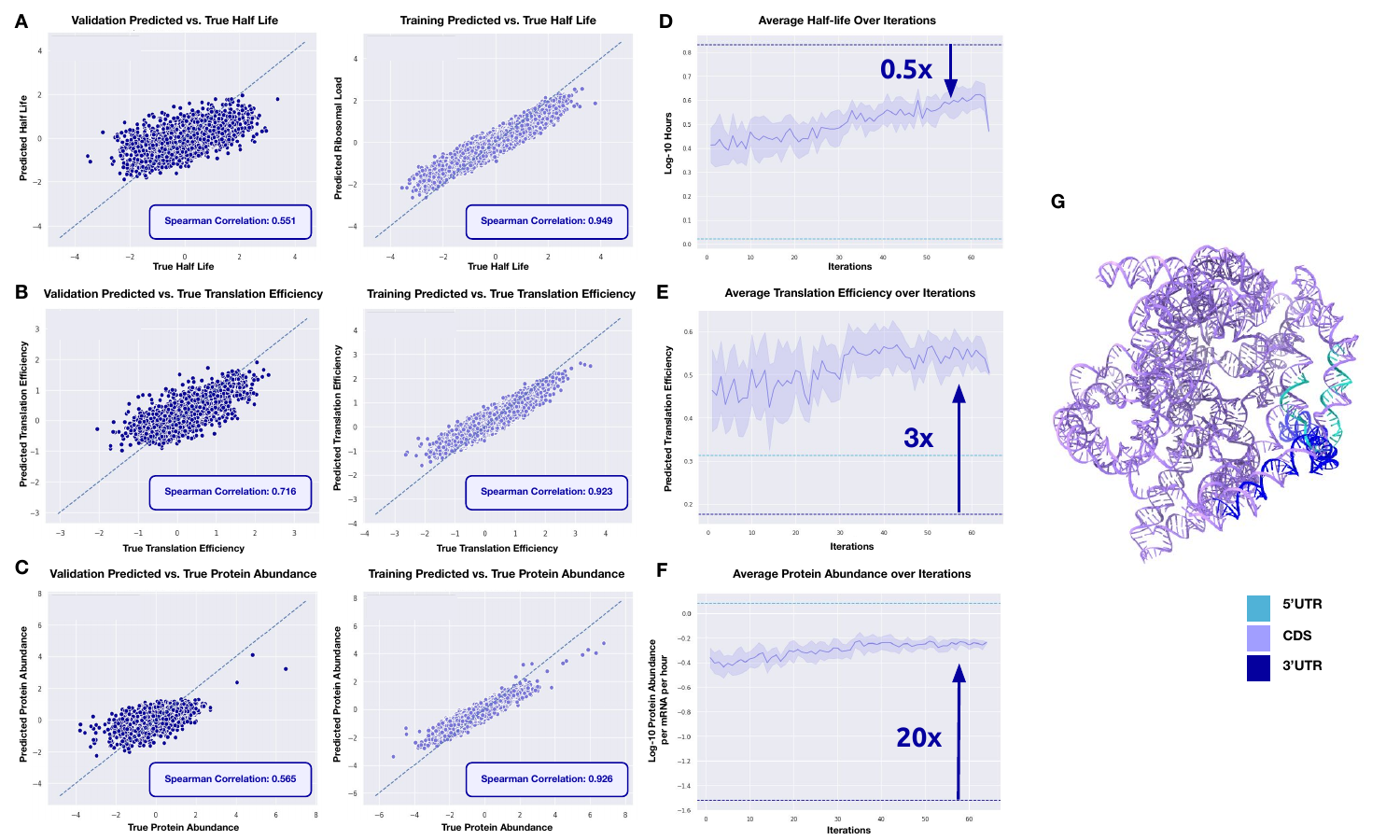}
    \vspace{-10pt}
    \caption{\textbf{Correlation plots for property regressors.} \textbf{(A)} Half-life, \textbf{(B)} translation efficiency, and \textbf{(C)} protein abundance correlation for validation (left) and training (right) data. Optimization curves for \textbf{(D)} half-life, \textbf{(E)} translation efficiency, and \textbf{(F)} protein abundance for conditional generation of \textit{P. pyralis} luciferase mRNA via Monte Carlo Tree Guidance (MCTG). Shown are Pareto front scores with standard error (purple), WT \textit{P. pyralis} luciferase score (navy), and median classifier property score from the entire dataset (teal).  \textbf{(G)} Predicted structure for a single generated \textit{P. pyralis} luciferase mRNA colored by sequence component (5'UTR: teal, CDS: purple, 3'UTR: navy).}
    \label{fig:properties}
\end{figure}

\subsection{Pretraining a Masked Discrete Diffusion Model on Full-Length mRNAs}

Effective mRNA design requires a generative model that can represent full transcripts faithfully, and that can be readily tuned toward external reward signals. Masked discrete diffusion models (MDMs) are a natural fit: they enable flexible, any-order generation from partially or fully masked sequences, capture bidirectional dependencies across long contexts, and have proven amenable to property guidance in both protein \citep{gruver2023protein} and peptide \citep{tang2025peptune} design. We therefore trained an MDM on 14.2 million vertebrate mRNA sequences from the Ensembl corpus \citep{dyer2025ensembl}, using a hybrid tokenization scheme that separately encodes CDSs (by codon) and UTRs (by single nucleotide) to enable targeted design of each region (see Section~\ref{data-collection}). The resulting model, mRNAutilus, parameterizes a denoiser over the joint UTR--CDS--UTR vocabulary (Figure~\ref{fig:overview}A).

To verify that mRNAutilus had learned a faithful distribution of natural mRNAs before applying any guidance, we generated three libraries of 200 sequences at token lengths 1,000, 2,500, and 5,000 via fully-masked ancestral sampling, and compared each against length-matched vertebrate mRNAs from the Ensembl corpus \citep{dyer2025ensembl}. Because our hybrid tokenization admits variable coding-sequence lengths, we normalized all length-dependent metrics for fair comparison. mRNAutilus-generated mRNAs exhibited slightly elevated GC content relative to natural mRNAs, converging toward the natural distribution at longer lengths (Figure~\ref{fig:unconditional}A), and lower length-normalized minimum free energies (MFEs) across all lengths as predicted by ViennaRNA \citep{lorenz2011viennarna}, indicating greater thermodynamic stability \citep{zhang2024deep} (Figure~\ref{fig:unconditional}B). The frequency of Kozak consensus motifs \citep{kozak1999initiation, kozak1987analysis, kozak1991analysis}, computed via the regular expression \texttt{(A|G)CCATGG}, closely matched natural frequencies (Figure~\ref{fig:unconditional}C), while pairwise sequence diversity (Figure~\ref{fig:unconditional}D) and per-sequence Shannon entropy (Figure~\ref{fig:unconditional}E) were nearly identical to those of evolutionarily selected Ensembl sequences, confirming that mRNAutilus does not collapse into the repetitive-token failure modes common to generative language models. Together, these results establish that mRNAutilus generates full-length mRNAs that recapitulate the statistical hallmarks of natural transcripts, providing a reliable foundation for downstream property guidance.

\subsection{mRNAutilus Embeddings Enable Functional Property Prediction from Sequence}

Guided generation requires reward signals that score candidate sequences for therapeutic fitness. Because mRNAutilus is a BERT-style MDM, we hypothesized that its internal representations should encode both coding structure and functional context, properties that could be exploited by lightweight regressors without retraining the backbone. We first confirmed that mRNAutilus embeddings discriminate coding from non-coding RNAs: projecting token-averaged embeddings for 100 mRNAs and 100 non-coding RNAs (ncRNAs) onto their first two principal components cleanly separated the two populations (Figure~\ref{fig:pca}A). We then trained XGBoost regressors \citep{chen2016xgboost} on mRNAutilus embeddings to predict three functional properties central to mRNA fitness: half-life \citep{agarwal2022genetic}, which modulates protein-product abundance and dosing; translation efficiency \citep{zheng2025predicting}, which captures ribosome recruitment as measured by polysome profiling; and protein abundance \citep{eichhorn2014mrna}, the ultimate functional readout of an mRNA therapeutic. Hyperparameters are provided in Table~\ref{tab:xgboost_params_regression}.

All three regressors achieved moderate-to-high correlation with ground truth on held-out validation sets (Figure~\ref{fig:properties}A-C), and XGBoost consistently outperformed KNN alternatives \citep{nguyen2024optimized, guo2022active} on these mRNA-specific tasks (Table~\ref{table:knn}). We then benchmarked mRNAutilus embeddings against other nucleic acid language models proposed for (m)RNA design \citep{wood2025helix, brixi2026genome, penic2025rinalmo, nguyen2023hyenadna}. mRNAutilus achieved the highest validation $R^2$ on half-life prediction and remained competitive on translation efficiency and protein abundance despite having an order of magnitude fewer parameters than Evo-2 (Table~\ref{tab:benchmarking}. These three regressors, combined with mRNAutilus itself, supply the reward vector needed to steer generation toward jointly optimal therapeutic properties.

\subsection{Multi-Objective Guided Generation Produces Functionally Superior mRNAs \textit{In Silico}}

Having established a generator and a set of embedding-based property predictors, we combined them via Monte-Carlo Tree Guidance (MCTG), following the multi-objective guidance strategy previously validated on peptide design by PepTune \citep{tang2025peptune}. MCTG couples the pre-trained MDM with the regressor-derived reward vector to navigate the vast mRNA design space toward Pareto-optimal sequences under half-life, translation efficiency, and protein abundance. We designed transcripts encoding three targets of engineering and therapeutic relevance: \textit{P. pyralis} luciferase (Fluc), SARS-CoV-2 Spike glycoprotein (S-Protein), and human mucin 1 (MUC1). Notably, neither luciferase nor spike was present in the mRNAutilus pretraining or regressor-training corpora, as they derive from invertebrate and viral sources, respectively, providing a stringent out-of-distribution test.

Simultaneous codon optimization and UTR design yielded transcripts with enhanced predicted half-life, translation efficiency, and protein abundance across all three targets, with the sole exception of predicted protein abundance for MUC1 (Table~\ref{table:Results}). Across MCTG iterations, average property values of rolled-out sequences increased consistently along almost all objectives, confirming effective multi-property guidance (Figure~\ref{fig:properties}D for Fluc; analogous curves for S-Protein and MUC1 in Figure~\ref{fig:cond-extra}). Characterizing the Fluc library further, we found that the generated 5' and 3'UTRs were slightly more thermodynamically stable and more GC-rich than natural vertebrate UTRs (Figure~\ref{fig:utrs}A,B) -- both hallmarks of improved translation efficiency \citep{metkar2024tailor, sample2019human, mauger2019mrna} -- yet highly diverse at the level of predicted secondary structure, indicating minimal structural homology across the library (Figure~\ref{fig:utrs}C). An ablation over the number of MCTG iterations and Pareto-front size confirmed robustness across hyperparameters (Figure~\ref{fig:ablation}). 

\begin{figure*}[h!]
    \centering
    \includegraphics[width=\linewidth]{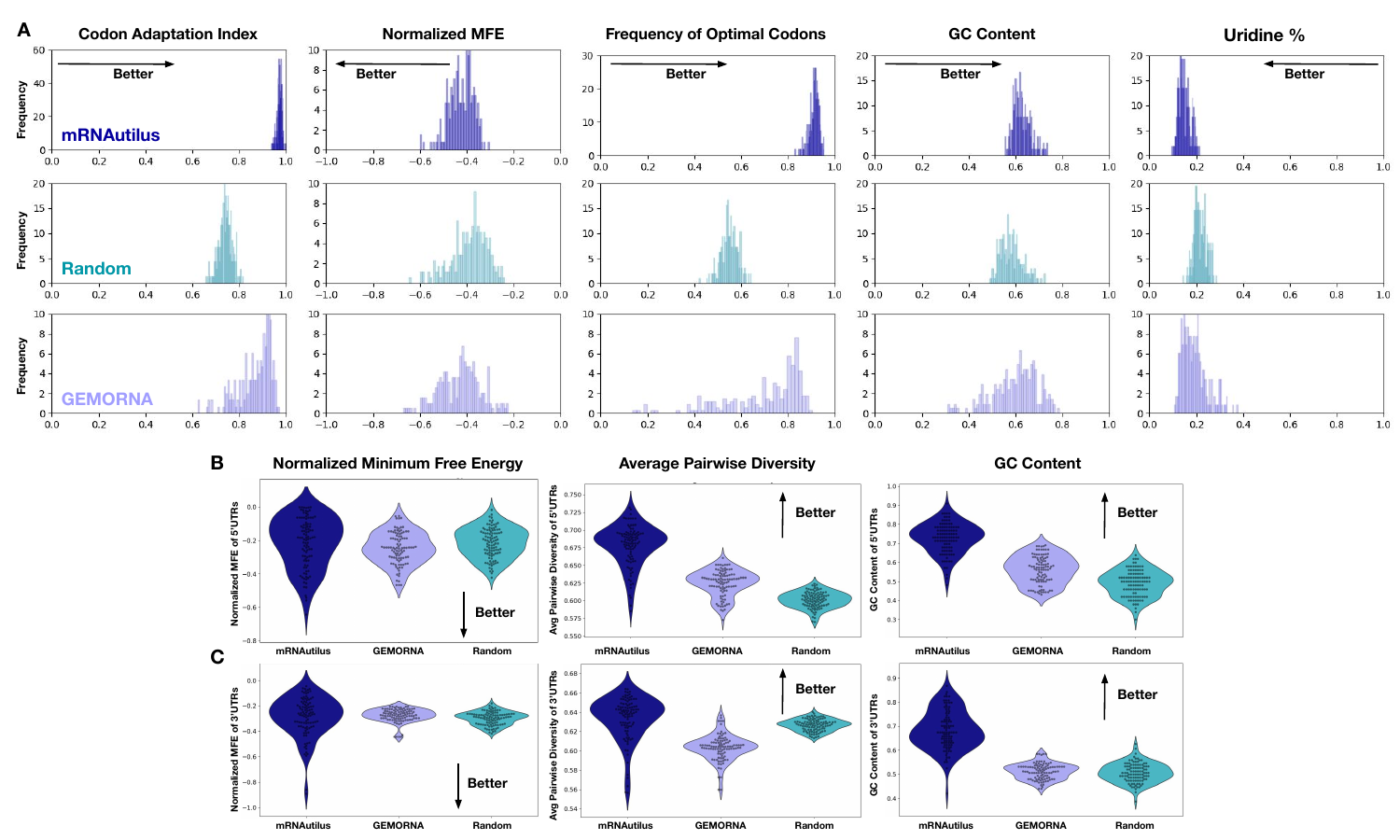}
    \caption{\textbf{Metric-driven evaluation of \textit{in-silico} full-sequence mRNA generation methods.} A library of 250 luciferase mRNAs was produced using mRNAutilus, randomness, and GEMORNA. All three methods were prompted to design a luciferase mRNA encoding the GenScript F-Luc protein. \textbf{(A)} Coding sequence comparison using all three methods. Considered metrics are codon adaptation index (CAI), length-normalized minimum-free energy (MFE), optimal codon frequency, GC content, and uridine percentage. Metric distributions are also shown for all generated \textbf{(B)} 5'UTRs and \textbf{(C)} 3'UTRs.}
    \label{fig:gemorna-comp}
\end{figure*}

We next benchmarked mRNAutilus directly against GEMORNA \citep{zhang2025deep}, a component-wise autoregressive mRNA design method, and against a random-design baseline generated by sampling each UTR nucleotide uniformly from $\{\text{A}, \text{C}, \text{G}, \text{T}\}$ and each codon uniformly from the synonymous-codon alphabet. Across 200-sequence libraries targeting the GenScript F-Luc protein, mRNAutilus-designed CDSs achieved superior codon adaptation index (CAI), length-normalized MFE, optimal-codon frequency, and uridine percentage (Figure~\ref{fig:gemorna-comp}A), and produced more thermodynamically stable and diverse UTRs than either comparator (Figure~\ref{fig:gemorna-comp}B,C). The less-negative tail of the mRNAutilus UTR MFE distribution has been linked to improved translation efficiency in certain contexts \citep{mauger2019mrna}, suggesting that this exploration reflects productive diversification of the UTR design space rather than instability.

\subsection{mRNAutilus-Designed mRNAs Achieve Superior Expression and Stability \textit{In Vitro}}

\begin{figure*}[h!]
    \centering
    \includegraphics[width=\linewidth]{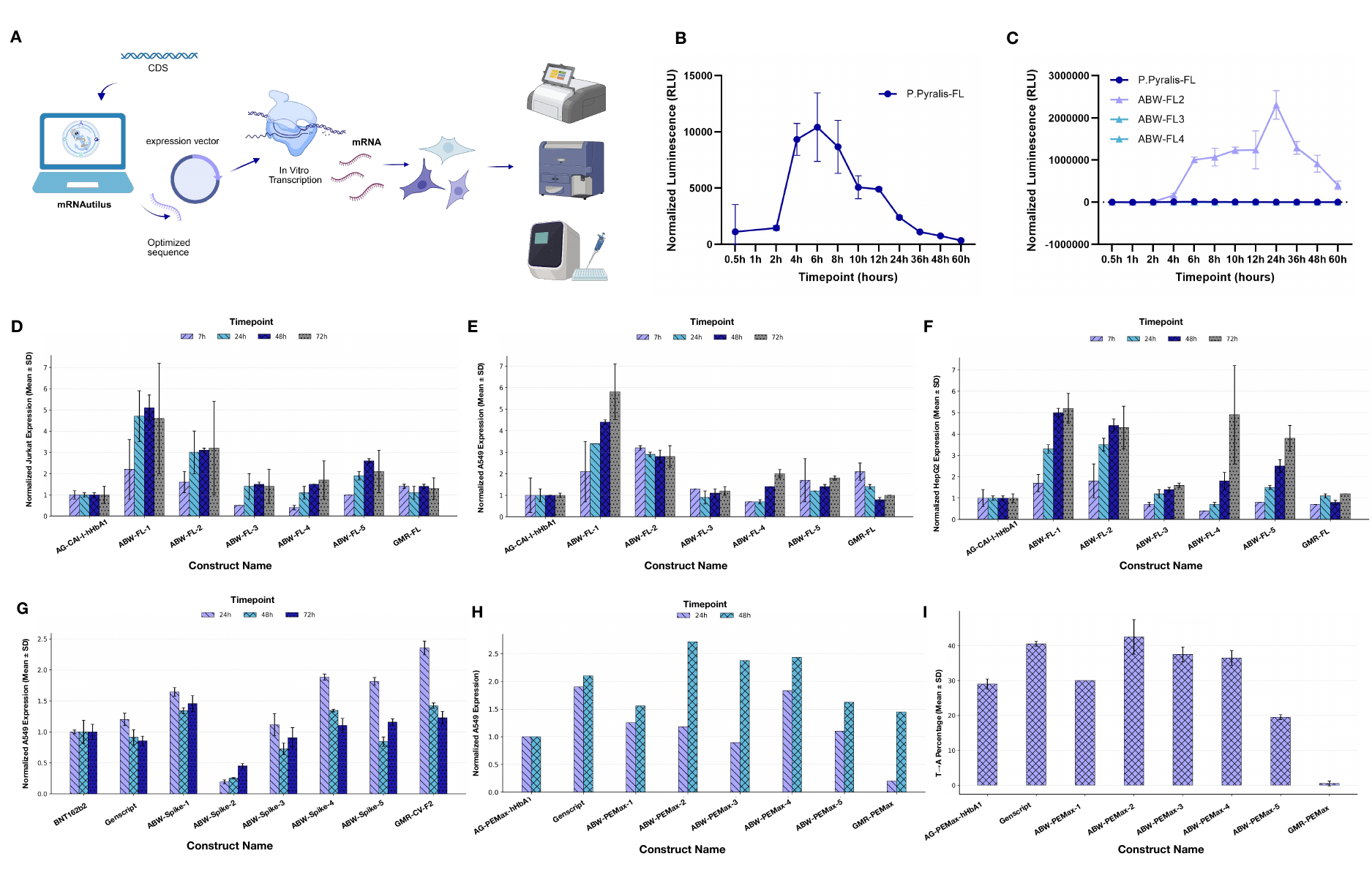}
    \vspace{-10pt}
    \caption{\textbf{mRNAs designed by mRNAutilus achieve superior zero-shot performance over competing methods \textit{in vitro}.}
    \textbf{(A)} Schematic of the experimental validation workflow for mRNAutilus-designed sequences. \textbf{(B)} Absolute expression levels for \textit{P. pyralis} luciferase mRNAs. \textbf{(C)} Normalized expression levels of \textit{P. pyralis} luciferase mRNAs over time in HEK293T cells (n=3 biological replicates). Error bars denote the standard deviation from the mean across replicates. Sequences produced by mRNAutilus (abbreviated as ABW-) are compared to WT \textit{P. pyralis} luciferaese mRNA. Normalized expressions for \textit{P. pyralis} luciferase mRNAs were then compared to a zero-shot GEMORNA-designed luciferase mRNA and a GenScript F-Luc mRNA coding sequence paired with human alpha-globin UTRs in \textbf{(D)} Jurkat, \textbf{(E)} A549, and \textbf{(F)} HepG2 cells. \textbf{(G)} Zero-shot mRNAs were generated for SARS-CoV-2 Spike glycoprotein using mRNAutilus, which were compared alongside several benchmarks. Displayed are normalized expressions in A549 cells. All mRNAs encode the same Spike glycoprotein antigen. \textbf{(H)} PEMax zero-shot mRNAs were designed using mRNAutilus alongside benchmark sequences in an expression assay within A549 cells. All mRNAs encode the same PEMax enzyme. \textbf{(I)} PEMax mRNAs were evaluated for editing efficiency at the HEK3 locus in A549 cells, where we evaluate the T$\rightarrow$A edit percentage to quantify functional efficacy.}
    \label{fig:Fluc}
\end{figure*}

Having demonstrated strong \textit{in silico} performance, we next tested whether mRNAutilus-designed sequences translate into functional gains in cells. We selected three targets with distinct biological and therapeutic relevance -- \textit{P. pyralis} luciferase, SARS-CoV-2 Spike glycoprotein, and PEMax (a prime-editor payload) -- and for each, drew candidate sequences from libraries jointly optimized for half-life and translation efficiency, benchmarking against a human alpha-globin (HAB) UTR composition as an \textit{in silico} baseline (Table~\ref{table:invitro}). For each target, the top five designs by predicted fold-change over the HAB baseline were synthesized by \textit{in vitro} transcription and evaluated in cell-based expression assays (Figure~\ref{fig:Fluc}A), following the protocols detailed in Section~\ref{sec:methods}.
For luciferase, these zero-shot designs produced luminescence signals comparable to or significantly greater than the wild-type control in HEK293T cells, with ABW2 exhibiting approximately 400-fold higher expression than wild-type \textit{P. pyralis} luciferase at 48~h post-transfection (Figure~\ref{fig:Fluc}B,C). Extending across three additional cell lines, the same mRNAutilus designs outperformed both a zero-shot GEMORNA-designed luciferase mRNA (GMR-FL) and a composition combining the GenScript F-Luc CDS with human alpha-globin UTRs in Jurkat (Figure~\ref{fig:Fluc}D), A549 (Figure~\ref{fig:Fluc}E), and HepG2 (Figure~\ref{fig:Fluc}F) cells. We then applied the same design procedure to SARS-CoV-2 Spike glycoprotein, using the GenScript Spike mRNA v2 ORF as the template. Three of the four designs (ABW-Spike-1, ABW-Spike-2, ABW-Spike-3) expressed above both BNT162b2 and the commercial GenScript Spike v2 mRNA in A549 cells, matched the expression profile of the lab-in-the-loop multi-shot GEMORNA sequence GMR-CV-F2 \citep{zhang2025deep} across all measured time points, and in the case of ABW-Spike-1, exhibited improved intracellular stability (Figure~\ref{fig:Fluc}G). 

We further extended our evaluation to PEMax, an optimized prime editor that enables precise, programmable genome edits (substitutions, insertions, and deletions) without requiring double-strand breaks~\citep{chen2021enhanced}. As a large fusion of a nickase Cas9, a codon-optimized reverse transcriptase, and multiple nuclear localization signals, PEMax presents a demanding test case for mRNA design, with a long coding sequence and complex domain architecture relative to compact reporters. Several mRNAutilus-designed constructs show expression comparable to or exceeding the commercial GenScript PEMax mRNA, with the advantage becoming more pronounced at 48h (Figure~\ref{fig:Fluc}H). ABW-PEMax-2 achieves the highest editing efficiency of all tested constructs, surpassing GenScript PEMax in T$\rightarrow$A conversion (Figure~\ref{fig:Fluc}I), with ABW-PEMax-3 and ABW-PEMax-4 also editing at comparable rates. In contrast, our zero-shot PEMax sequences express and edit an order of magnitude higher than the GEMORNA zero-shot PEMax, which shows negligible editing activity. Together, these results demonstrate that zero-shot mRNAutilus generation produces mRNAs with superior expression and durability across reporter, antigen, and genome editing payloads without any lab-in-the-loop iteration.

\subsection{mRNAutilus Enables Functional Intracellular Protein Degradation via Optimized Degrader mRNAs}

\begin{figure*}[t]
    \centering
    \includegraphics[width=\linewidth]{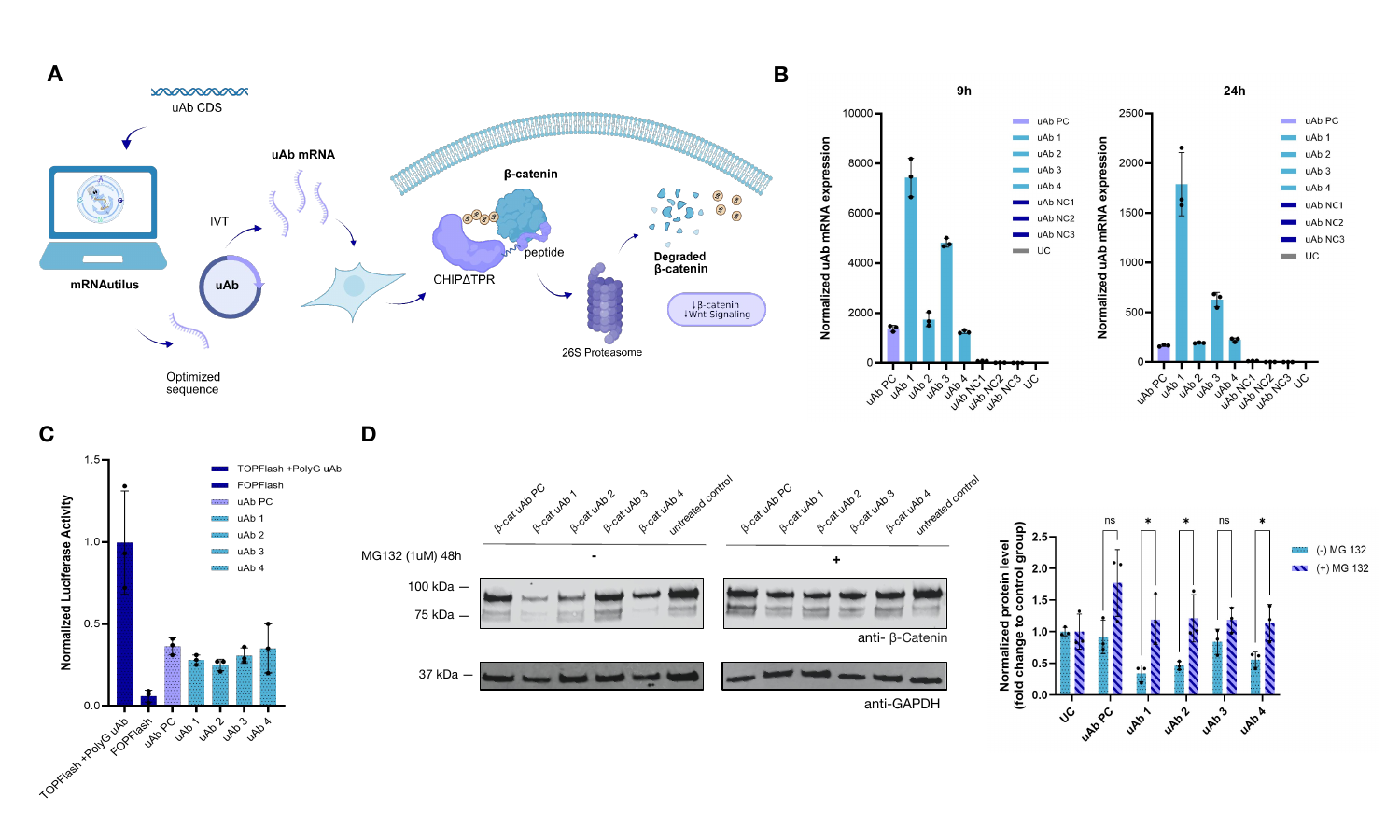}
    \vspace{-10pt}
    \caption{\textbf{Design of therapeutic mRNAs with mRNAutilus and \textit{in vitro} evaluation.}  \textbf{(A)} Architecture and mechanism of the uAb degradation system with sequences optimized by mRNAutilus. CHIP$\Delta$TPR fused to the C-terminus of $\beta$-catenin-specific peptides targets endogenous proteins for ubiquitin-mediated degradation via the proteasome following mRNA transfection. Created with BioRender. \textbf{(B)} Quantitative PCR analysis of uAb mRNA expression levels. Cells were harvested at 9~h and 24~h post-transfection. Values were normalized to the untreated control. \textbf{(C)} TOPFlash luciferase reporter assay of Wnt/$\beta$-catenin transcriptional activity. The FOPFlash reporter served as a negative control. All values were normalized to the TOPFlash + polyG uAb control condition. \textbf{(D)} Left: Degradation of endogenous $\beta$-catenin in cytosolic fractions of DLD-1 cells analyzed by immunoblotting using anti-$\beta$-catenin and anti-GAPDH antibodies. Right: Densitometric analysis of immunoblots ($n = 3$) was performed using ImageJ to quantify $\beta$-catenin levels. Statistical analysis for each condition was performed using a two-tailed unpaired Student's $t$-test. Data are presented as mean $\pm$ SD, and calculated $p$ values are represented as follows: *$p \leq 0.05$; ns, not significant.}
    \label{fig:functional}
\end{figure*}

To probe whether mRNAutilus generalizes beyond reporters and antigens to functionally demanding intracellular payloads, we turned to ubiquibodies (uAbs): synthetic E3 ligases in which the native substrate-binding domain of CHIP is replaced with a language model-designed peptide binder, redirecting the endogenous ubiquitin-proteasome machinery to degrade a target of choice \citep{portnoff2014ubiquibodies, brixi2023salt, chen2025target, bhat2025novo, ye2024programmable}. uAbs are particularly attractive as mRNA payloads because mRNA-LNP delivery enables transient, tunable intracellular expression of the degrader without genomic integration. We targeted $\beta$-catenin, a driver of Wnt-pathway oncogenesis whose stabilization promotes \textit{c-Myc} and \textit{cyclin D-1} expression, using an existing uAb previously validated in our PepPrCLIP work \citep{bhat2025novo} (Figure ~\ref{fig:functional}A). Four mRNAutilus-designed mRNAs with the highest predicted fold-change over the HAB baseline were synthesized; upon transfection, the encoded uAb is expressed intracellularly, binds $\beta$-catenin via its peptide-binding domain, and recruits CHIP$\Delta$TPR to drive proteasomal degradation (Figure~\ref{fig:functional}A).

We first assessed transcript abundance by qPCR at 9 and 24~h post-transfection in DLD-1 cells. At both time points, optimized uAb mRNAs exhibited markedly higher expression than the human $\alpha$-globin-based positive control, while deoptimized negative controls showed poor expression (Figure~\ref{fig:functional}B), confirming that mRNAutilus sequence optimization translates to substantially improved transcript-level output even for this complex payload. We next asked whether elevated expression preserved (or enhanced) downstream degrader function. In the TOPFlash Wnt/$\beta$-catenin reporter assay, optimized uAb mRNAs fully retained their intended activity, suppressing Wnt/$\beta$-catenin transcriptional signaling relative to the TOPFlash + polyG uAb control, with some designs showing modestly greater suppression than the positive control (Figure~\ref{fig:functional}C). Immunoblot analysis of cytosolic fractions confirmed this functional engagement at the protein level: optimized constructs uAb1, uAb2, and uAb4 showed significant reductions in endogenous $\beta$-catenin by densitometry, and these reductions were abolished upon co-treatment with the proteasome inhibitor MG132, confirming that the observed loss of $\beta$-catenin is mediated through the ubiquitin--proteasome pathway rather than off-target effects (Figure~\ref{fig:functional}D). Together, these findings demonstrate that mRNAutilus optimization improves both expression and functional efficacy of uAb mRNAs, enabling potent proteasome-dependent degradation of endogenous disease-relevant targets and establishing the framework as broadly applicable to programmable intracellular therapeutic modalities.

\section{Discussion}
The rational design of therapeutic mRNAs requires jointly optimizing multiple competing properties -- stability, translation efficiency, and expression -- a challenge that traditional single-objective approaches struggle to address systematically. Here we introduce mRNAutilus, to our knowledge the first multi-objective-guided generative model for full-length mRNA design, combining a masked discrete diffusion backbone with Monte Carlo Tree Guidance (MCTG) \citep{tang2025peptune} to produce Pareto-optimal sequences under simultaneous codon and UTR design. Unconditional generation recapitulates the statistical hallmarks of natural mRNAs, while lightweight XGBoost regressors over mRNAutilus embeddings provide reliable reward signals for MCTG-guided sampling. Critically, we demonstrate improved \textit{in vitro} performance for protein-coding genes outside the mRNAutilus training distribution (\textit{P. pyralis} luciferase, SARS-CoV-2 Spike, PEMax gene-editor, and a designer $\beta$-catenin uAb), indicating that the framework generalizes beyond its pretraining corpus. This approach is gradient-free and remains viable in low-data regimes where regressor training sets fall below $\sim$100 points, as is typical for emerging experimental mRNA datasets.

Several directions promise to extend these capabilities. On the algorithmic front, recent advances in guidance for discrete generative models -- trajectory-aware tree search for discrete diffusion (TR2-D2 \citep{tang2025tr2}) and multi-objective guidance for discrete flow matching (MOG-DFM, AReUReDI \citep{chen2025mogdfm, chen2025areuredi}) -- offer complementary machinery that could be integrated into the mRNAutilus pipeline. A more immediate limitation, however, lies upstream of the model: property guidance is only as reliable as the regressors driving it, and public mRNA datasets remain sparse, cell-type-restricted, and heterogeneous in their readouts \citep{agarwal2022genetic, zheng2025predicting, eichhorn2014mrna}. Addressing this will require close collaboration with experimentalists to generate large, uniformly measured datasets spanning half-life, translation efficiency, protein abundance, immunogenicity, and tissue-specific expression, ideally embedded in a design-build-test-learn loop that refines both regressors and generator.

Looking ahead, we envision mRNAutilus as a general-purpose method for programmable mRNA therapeutic design. The demonstrated capacity to optimize reporter proteins, viral antigens, genome editing machinery (PEMax), and peptide-guided E3 ligases suggests broad applicability across vaccines, protein replacement therapies, prime and base editing, and programmable proteome modulation via uAbs and related proteome editing architectures~\citep{hong2025duab, hong2026pepreps}. Coupled with lipid nanoparticle delivery, mRNAutilus-designed transcripts enable transient, tunable, and target-specific protein expression, offering a template for how pretrained discrete diffusion paired with multi-objective guidance can transform sequence-based therapeutic design across biomolecular modalities.

\section{Materials and Methods}
\label{sec:methods}
\subsection{Masked Discrete Diffusion Framework}
\subsubsection{Discrete Diffusion for Sequence Generation}
Let $\mathbf{x}=(\mathbf{x}^1, \dots, \mathbf{x}^L) \in \mathcal{V}^L$ denote a sequence with length $L$ over a vocabulary $\mathcal{V}$ of $V$ discrete states. For notational simplicity, we develop the masked discrete diffusion formulation for a single token $\mathbf{x}\in \Delta^{V-1}$, where $\Delta^{V-1}$ denotes the probability simplex over $V$ states. The extension to a sequence of length $L$ follows from factorization.

We denote $p_0(\mathbf{x})=p_{\text{data}}(\mathbf{x})$ as the target data distribution,  $p_1(\mathbf{x})=p_{\text{prior}}(\mathbf{x})$ as the prior (noisy) distribution, and $p_t(\mathbf{x})$ as the time-dependent marginal distribution for $t \in [0,1]$ that interpolates between $p_0$ at $t=0$ and \(p_1\) at \(t=1\). Given a clean sequence $\mathbf{x}_0\sim p_0$, represented as a one-hot vector $\mathbf{x}_0\in \{0,1\}^V$, and a noisy sequence $\mathbf{x}_1\sim p_1$, the \textit{reverse posterior} as the conditional probability of obtaining $\mathbf{x}_0$ given $\mathbf{x}_t$ as $p_{0|1}(\mathbf{x}_0|\mathbf{x}_1)$. 

The discrete diffusion framework \citep{austin2021structured} enables the generation of sequences from the reverse posterior via a two-stage training algorithm. The forward noising process applies categorical noise to $\mathbf{x}_0$ over continuous time steps $t\in [0,1]$ according to:
\begin{align}
    p_{t|0}(\mathbf{x}_t|\mathbf{x}_0)=\mathrm{Cat}\left( \mathbf{x}_t;\, \alpha_t \mathbf{x}_0 + (1-\alpha_t)\,\boldsymbol{\pi} \right)
\end{align}
where $\boldsymbol{\pi}\in \Delta^{V-1}$ is a fixed noise distribution and $\alpha_t$ is a time-dependent schedule specifying the probability that a token remains unchanged. The marginal $p_t$ is then defined as the expectation of conditional distributions:
\begin{align}
    p_t(x_t) = \mathbb{E}_{p_{\mathrm{data}}}\left[ p_{t|0}(\mathbf{x}_t | \mathbf{x}_0) \right]\label{eq:gen_flow_background}
\end{align}
which yields a tractable learning objective by conditioning on training sequences in the dataset $x_0\sim p_{\text{data}}$.

The backward denoising process iteratively recovers the clean sequence from noise using the marginal reverse posterior $p_{s|t}(\mathbf{x}_s|\mathbf{x}_t)$ that defines the probability over slightly less noisy sequences $\mathbf{x}_s$ at time $s=t-\frac{1}{T}$, where $T$ is the total number of sampling steps. Given a clean sequence $\mathbf{x}_0$, the \textit{conditional reverse posterior} is defined as $p_{s|t, 0}(\mathbf{x}_s|\mathbf{x}_t, \mathbf{x}_0)$. We train a neural network with parameters $\theta$ to predict the distribution $p_{s|t, 0}(\mathbf{x}_s|\mathbf{x}_t, \mathbf{x}_0)$ at training time, and sample from the parameterized reverse posterior $p_{s|t}^\theta(\mathbf{x}_s|\mathbf{x}_t)$ at inference.

\subsubsection{Masked Discrete Diffusion Models}
We adopt the masked variant of discrete diffusion for its demonstrated efficacy in generating high-quality sequences with long-range dependencies \citep{Sahoo2024, Shi2024, ou2024, Zheng2024}. Under this formulation, the forward noising process applies an absorbing mask token via:
\begin{equation}
\label{eq:mask_flow}
    p_{t|0}^{\mathrm{mask}}(\mathbf{x}_t | \mathbf{x}_0) = \mathrm{Cat}\left( \mathbf{x}_t; t\mathbf{x}_0+(1-t) \mathbf{m}\right)
\end{equation}
where \(\mathbf{m}\) denotes the one-hot vector of the [MASK] token. At $t=1$, the distribution converges to only the fully masked sequence $p_{t|0}^{\mathrm{mask}}(\mathbf{x}_t | \mathbf{x}_0) =\delta(\mathbf{x}_t-\mathbf{m})$, and at \(t=0\), the distribution converges to the clean sequence $p_{t|0}^{\mathrm{mask}}(\mathbf{x}_t | \mathbf{x}_0) =\delta(\mathbf{x}_t-\mathbf{x}_0)$. Applying Bayes' rule yields the conditional reverse posterior:
\begin{align}
    p_{s|t, 0}(\mathbf{x}_s|\mathbf{x}_t, \mathbf{x}_0)=\begin{cases}
        \left(1-\frac{s}{t}\right)\mathbf{x}_0+\frac{s}{t}\mathbf{m}&\mathbf{x}_t=\mathbf{m}\\
        \mathbf{x}_s &\mathbf{x}_s\neq \mathbf{m}
    \end{cases}
\end{align}
which means that once a token is unmasked, it remains fixed for the remainder of the reverse process. Since the clean sequence $\mathbf{x}_0$ is unknown at inference, we train a parameterized denoiser $p^\theta_{s|t}(\mathbf{x}_s|\mathbf{x}_t)$ with parameters $\theta$ to predict the conditional reverse posteriors from $\mathbf{x}_t$ alone. The procedure is outlined in Figure ~\ref{fig:overview}A.

\subsubsection{Training Objective}
We approximate the denoising distribution $p_{0|t}(\mathbf{x}_0 | \mathbf{x}_t)$ with a neural network $p^\theta_{0|t}(\mathbf{x}_0 | \mathbf{x}_t)$. During training, each batch is corrupted with mask tokens at uniformly sampled time steps $t\sim \mathcal{U}(0,1)$, and the model predicts $\mathbf{x}_0^\theta\in \Delta^{V-1}$, which is a categorical distribution over the vocabulary indicating the predicted probability that the clean token exists in each discrete state. The model parameters are optimized by minimizing the continuous-time negative evidence lower bound (NELBO) between the predicted distribution $x_0^\theta$ and the clean one-hot vector $\mathbf{x}_0$ given by:
\begin{equation}
\label{eq:ce_loss_background}
\mathcal{L}(\theta) = \mathbb{E}_{t, p_{t|0}(\mathbf{x}_t|\mathbf{x}_0), p_0(\mathbf{x}_0)} \left[ -\frac{1}{t}\log \left\langle \mathbf{x}_0, \mathbf{x}_0^\theta\right\rangle \right]
\end{equation}
following the continuous-time formulation from \citep{Sahoo2024}.  This objective is equivalent to a time-weighted negative log-likelihood and vanishes when $\mathbf{x}_0 =\mathbf{x}^\theta_0$. 

\subsubsection{Unconditional Sampling}
To sample from the learned data distribution, we initialize a fully masked sequence $x_1=[M]^L$ of length $L$ and take $T$ discretized reverse-time steps $t=\{1, \dots, \frac{1}{T}\}$. At each step, we evaluate the denoiser to obtain $\mathbf{x}_0^\theta\approx p_{0|t}^\theta(\mathbf{x}_0|\mathbf{x}_t)$ and compute the marginal reverse posterior:
\begin{align}
    p^\theta_{s|t}(\mathbf{x}_s|\mathbf{x}_t)=\begin{cases}
        \left(1-\frac{s}{t}\right)\mathbf{x}_0^\theta+\frac{s}{t}\mathbf{m}&\mathbf{x}_t=\mathbf{m}\\
        \mathbf{x}_s &\mathbf{x}_s\neq \mathbf{m}
    \end{cases}
\end{align}
To inject stochasticity, we apply Gumbel noise $g\sim -\log(-\log (u+\epsilon)+\epsilon)$, where $u\sim \mathcal{U}(0,1)$ and $\epsilon =10^{-10}$, and sample the argmax token from the perturbed distribution:
\begin{align}
    \mathbf{x}_s= \arg\max(\log p^\theta_{s|t}(\mathbf{x}_s|\mathbf{x}_t)+g)
\end{align}
Given our constraint that masked tokens remain unchanged after unmasking, we update $\mathbf{x}_s$ only when $\mathbf{x}_t=\mathbf{m}$. After $T$ unmasking steps, we ensure all tokens are unmasked and return the generated sequence $\mathbf{x}_0$.

\subsection{Pre-Trained mRNA Masked Discrete Diffusion Model}
\subsubsection{Encoder Architecture}
mRNAutilus is a BERT-style generative language model trained on a masked diffusion objective for mRNA sequence design, which parametrizes $p^\theta_{0|t}(\mathbf{x}_0|\mathbf{x}_t)$ defined above. Given an input sequence $\mathbf{x} = (\mathbf{x}^1, ..., \mathbf{x}^L)$ of length $L$, the model maps it to a sequence of token embeddings of shape $(L \times d)$. The architecture consists of several transformer blocks, where each block contains a multi-head self-attention module with 20 attention heads, a feed-forward network (FFN) using the SwiGLU activation function \citep{shazeer2020glu}, and Layer Normalization within residual connections to stabilize training dynamics. mRNAutilus adopts Rotary Positional Embeddings (RoPE) \citep{su2024roformer} to encode both absolute and relative positional information, thereby improving its ability to model long-range dependencies in mRNA sequences. To enhance computational efficiency, we also utilize FlashAttention-2 \citep{dao2023flashattention2}. In total, mRNAutilus has 150M parameters across all modules.

\subsubsection{Data} \label{data-collection}
Ensembl is a genome browser containing the largest curated, annotated deposit of vertebrate mRNA \citep{dyer2025ensembl}. Using the Ensembl Perl API, we queried the database released in version 113 and collected protein-coding mRNA transcripts across 342 vertebrate species. Overall, we obtained a total of approximately 14.2 million sequences. Transcripts were filtered to only include sequences with a valid coding sequence length, appropriate start/stop codons, and no missing UTRs. Following this procedure, 5,526,848 sequences remained. An overview of the dataset is shown in Figure \ref{fig:pretraining}A,B. We additionally collect metrics for full-sequences and UTRs within the pretraining dataset in addition to the proportion of species represented in Figure ~\ref{fig:pretraining}C.

We leverage a hybrid tokenization scheme, encoding the coding sequences (CDS) and UTRs for each mRNA transcript separately. We tokenize the coding sequence as 3-mers, effectively by codon. UTRs are separately encoded at single-base resolution. Following this process, the 5'UTR tokens, CDS tokens, and 3'UTR tokens are concatenated in that order. For the CDS, we reference the standard codon table, which spans 64 codons, 1 primary start codon, and 3 stop codons. 

For the UTR sequences, all uracil (U) bases are replaced with thymine (T) to maintain consistency across representations, resulting in a primary UTR vocabulary of four nucleotides: 'A', 'C', 'T', and 'G'. Additionally, we include IUPAC degenerate nucleotide codes such as 'I', 'R', 'Y', 'K', 'M', 'S', 'W', 'B', 'D', 'H', 'V', 'N', and '-' to account for sequence ambiguities commonly found in biological datasets. We do not discard any sequences. Additionally, for all generated sequences, the Kozak motif ('GCCACC') is appended to the 5'UTR before token unmasking.

We include special tokens [CLS], [EOS], [MASK], [UNK], and [PAD], which denote sequence classification, end-of-sequence markers, masked tokens for pretraining, and padding, respectively. Sequences are truncated or padded to a total of 7,500 tokens for batched training. In total, our vocabulary spans 86 tokens.

\subsubsection{Pretraining}
Our model is trained for 100K updates from scratch, with a token batch size of 200K tokens per gradient step. We additionally utilize a batch sampler to construct batches using sequences of similar lengths. During training, tokens are sampled for masking uniformly at random at a rate modulated by the sampled timestep. Additionally, any codon token not coding for the same amino acid as the codon in the true mRNA sequence is rejected during denoising. We find that including this rule slightly improves empirical performance on the training objective (not shown).

\subsection{Property Regressors}
mRNA efficacy is highly dependent on factors that are largely influenced by sequence. We train XGBoost regressors for several key mRNA properties, including half-life, translation efficiency, and protein abundance, from mRNAutilus learned sequence embeddings. Parameters for the XGBoost regressors are listed in Table ~\ref{tab:xgboost_params_regression}.

\paragraph{Half-Life}
The half-life of a substance is the time required for it to decay to half of its original amount (or concentration within a solution). While more complex in the context of therapeutics, as half-life partially controls protein product abundance, dosage, and overall design flexibility, it is a critical factor to consider and tune when designing an mRNA therapeutic. For example, the mRNA concentration must be controlled to produce an antigen that can produce an immune response, but not induce a severe immunogenic effect. 

\paragraph{Translation Efficiency}
Following transcription, processing (5'capping, splicing, and 3'polyadenylation), and nuclear export, a mature mRNA transcript will interact with several proteins before ultimately forming a complex with ribosomal subunits to initiate translation. Several ribosomes bound to an mRNA transcript form a polysome, which serves as a proxy for mRNA fitness up until an actual protein product is formed. Polysome formation can be quantified by polysome profiling, an analog for how efficiently an mRNA can be recruited for protein translation. 

\paragraph{Protein Abundance}
Protein expression is the ultimate purpose of mRNA therapeutics. Following polypeptide release, folding, and trafficking, a mature protein product can perform its function within the cell. Quantifying the projected protein product formed within the cell also fundamentally determines the efficacy of the originating transcript. In a therapeutic context, a sufficient concentration of antigen must exist in circulation to elicit an immune response. 

\paragraph{Regressor Training}
For half-life prediction, we use an XGBoost boosted tree \citep{chen2016xgboost} regression model using mRNAutilus embeddings that predicts half-life directly from mRNA sequence (Table \ref{tab:xgboost_params_regression}). Paired data was collected from \citep{agarwal2022genetic}, which sourced transcriptome-wide half-life data from human samples across 16 publications and deposits. Overall, this dataset spans nearly 13,000 transcripts across 39 human samples. 

Similarly to half-life, we train an XGBoost regression model on ribosome profiling data collected from \citep{zheng2025predicting} spanning 1,282 human and 995 mouse datasets across several cell lines. This data serves as a proxy for \textit{translation efficiency}.

Lastly, we train a regression model that predicts protein abundance (log-10 scale) from conjoined mRNA sequencing and ribosomal profiling experiments conducted in human B2 and U2OS cells \citep{eichhorn2014mrna}. 

Lightweight regressors implemented through XGBoost boosted trees, and K-Nearest Neighbor (KNN) algorithms have been widely adopted due to their computational efficiency and robustness in low-data settings \citep{nguyen2024optimized, guo2022active}. We additionally trained KNN regressors to compare with the XGBoost regressors.

\subsection{Multi-Objective-Guided Generation} \label{peptune}
\subsubsection{Pareto Optimality} 
In multi-objective optimization, rather than a single optimal solution, there is a set of \textit{Pareto-optimal solutions}, where not one objective can be improved without sacrificing performance in another objective. Intuitively, this set \textit{maximizes} the trade-offs between solutions such that all objectives are exhausted to their limit. 

Given a vector of objective functions $\mathbf{f}(\mathbf{x}) =
[f_1(\mathbf{x}), \ldots, f_K(\mathbf{x})]$ where $f_k :\mathcal{V}^L \to \mathbb{R}$, suppose we aim to identify the set of solutions $x$ that \textit{maximizes} all $K$ objective functions. For two sequences $\mathbf{x}, \mathbf{x}'$, we say $\mathbf{x}$ \textit{dominates} $\mathbf{x}'$ if for all $K$ objectives, the reward for sequence $\mathbf{x}$ is better than or equal to the score for sequence $\mathbf{x}'$ and for at least one reward, the reward is strictly better:
\begin{align}
    \underbrace{\forall k\in \{1, \dots, K\}\;\;f_k(\mathbf{x}) \geq f_k(\mathbf{x}')\;\;\mathrm{and}\;\;\exists k^\star\;\;f_{k^\star}(\mathbf{x})> f_{k^\star}(\mathbf{x}')}_{\text{denoted as }\mathbf{f}(\mathbf{x})\succ \mathbf{f}(\mathbf{x}')}
    \tag{$\mathbf{x}$ dominates $\mathbf{x}'$}
\end{align}
A Pareto-optimal sequence $\mathbf{x}^\star$ is \textit{non-dominated}, such that no other sequence in the solution space $\mathbf{x}'\in\mathcal{X}$ \textit{dominates} it. This means it is just as optimal or more optimal than any other sequence in the solution space. Given that there are a potentially infinite number of trade-offs between objectives, we can define a \textit{Pareto-optimal set} $\mathcal{P}^\star$ where all sequences in the set are non-dominated.
\begin{align}
    \mathcal{P}^\star=\{\mathbf{x}\mid \nexists \mathbf{x}'\in \mathcal{X}, \;\mathbf{f}(\mathbf{x}')\succ\mathbf{f}(\mathbf{x})\}
\end{align}
Since the size of $\mathcal{P}^\star$ is potentially unbounded, the goal of multi-objective optimization is to approximate a finite set of Pareto-optimal solutions in a tractable manner.

\subsubsection{Monte-Carlo Tree Guidance}
To guide the unconditional MDM towards Pareto-optimality across multiple properties relevant to mRNA therapeutic efficacy, we leverage the Monte Carlo Tree Guidance (MCTG) algorithm introduced in \citet{tang2025peptune} for multi-objective guidance of discrete diffusion across several pre-trained regressor models (Algorithm \ref{alg:MCTS}). This is outlined in Figure ~\ref{fig:overview}B.

\subsubsection{Initialization} 
We initialize a tokenized sequence with a partially masked coding sequence (CDS), a masked 5'UTR sequence, and a masked 3'UTR sequence. This becomes the root node of the MCTS tree with timestep $t=1$. Additionally, we construct a vector scoring function $\mathbf{f} :\mathcal{V}^L \to \mathbb{R}^K$ that returns a score vector given a clean input sequence $\mathbf{x}_0 \in \mathcal{V}^L$ and initialize an empty Pareto-optimal set $\mathcal{P}^\star= \{\}$. We define the following hyperparameters: number of MCTS iterations $N_{\text{iter}}$ and number of children nodes $M$. A template CDS is also provided at input, which is referenced during expansion.

\subsubsection{Selection} 
To determine the next unmasking step from a parent node with an already expanded set of child sequences $\mathbf{x}_{\text{child}}^i\in \text{children}(\mathbf{x})$, we compute a selection score vector $\mathbf{U}(\mathbf{x}_{\text{child}}^i)$ defined as:
\begin{align}
    \mathbf{U}(\mathbf{x}_{\text{child}}^i)=\frac{\mathbf{W}(\mathbf{x}_{\text{child}}^i)}{N_{\text{visit}}(\mathbf{x}_{\text{child}}^i)}+c\cdot p^\theta_{s|t}(\mathbf{x}_{\text{child}}^i|\mathbf{x})\frac{\sqrt {N_{\text{visit}}(\mathbf{x})}}{1+N_{\text{visit}}(\mathbf{x}_{\text{child}}^i)}
\end{align}
which divides the cumulative reward vector $\mathbf{W}(\mathbf{x}_{\text{child}}^i)$ of a child node to the number of times it was visited $N_{\text{visit}}(\mathbf{x}_{\text{child}}^i)$ in addition to an exploration term with constant $c=0.1$. This encourages exploration of child nodes that have not been visited, in addition to high-reward nodes. Then, we select from the Pareto-optimal set of child nodes based on the normalized selection score vector. 
\begin{align}
    \mathcal{P}^\star_{\text{select}}=&\{\mathbf{x}_{\text{child}}^{i}\;|\;\nexists \mathbf{x}_{\text{child}}^j\in \text{children}(\mathbf{x})\quad \text{s.t.}\quad \mathbf{U}(\mathbf{x}_{\text{child}}^j)\succ \mathbf{U}(\mathbf{x}_{\text{child}}^i)\}
\end{align}
If we select a non-leaf node, we restart the selection process with the selected node as the new parent until we reach an expandable leaf node or a fully unmasked sequence. In the case of a fully unmasked sequence, we restart the selection from the root node. 

\subsubsection{Expansion} 
From the leaf node $\mathbf{x}_t$ at time $t$, we sample $M$ distinct sequences by applying Gumbel noise to the denoising distribution $p_{0|t}^{\theta}(\mathbf{x}_0|\mathbf{x}_t)$
\begin{align}
    \mathbf{x}_0^i\sim \log p_{0|t}^{\theta}(\mathbf{x}_0|\mathbf{x}_t)+\mathbf{g}_i
\end{align}
where $\mathbf{g}_i\sim \text{Gumbel}(0,1)\in \mathbb{R}^V$. Then, for each child node $\mathbf{x}_{\text{child}}^i$, we randomly select $k$ positions in $\mathbf{x}_t$ to unmask according to $\mathbf{x}_1^i$. 

During sampling, we employ constraints to ensure that an invalid mRNA sequence is not produced during generation. Specifically, we do not allow codon tokens to be sampled in the position of a UTR token. For any masked codon token, we permit only codons encoding for the same amino acid as the corresponding codon in the input template to be sampled. This constraint guarantees that the sampled mRNA sequence does not form an unintended protein product.

\subsubsection{Rollout} 
For each child node $\mathbf{x}_{\text{child}}^i$, we use self-planned ancestral sampling to obtain a fully unmasked sequence $\tilde{\mathbf{x}}_{\text{child}}^i$. Then, we feed each of the clean sequences into the vector score function $\mathbf{f}(\tilde{\mathbf{x}}_{\text{child}}^i)$ to obtain a score vector. We then compute the reward vector $\mathbf{r}(\mathbf{x}_{\text{child}}^i)\in \mathbb{R}^K$ of each child $\mathbf{x}_{\text{child}}^i$ where each element is the fraction of sequences in the Pareto-optimal set $\mathbf{x}^\star\in \mathcal{P}^\star$ that $\mathbf{x}_{\text{child}}^i$ dominates.
\begin{align}
    r_k(\mathbf{x}_{\text{child}}^i)=\frac{1}{|\mathcal{P}^\star|}\sum_{\mathbf{x}^\star\in \mathcal{P}^\star}\mathbf{1}\big[f_k(\mathbf{x}_{\text{child}}^i)\geq f_k(\mathbf{x}^\star)\big]
\end{align}
For all $M$ children sequences, we add $\mathbf{x}_{\text{child}}^i$ to $\mathcal{P}^\star$ if it is \textit{non-dominated} by all sequences $\mathbf{x}^\star\in \mathcal{P}^\star$ and remove all sequences in $\mathcal{P}^\star$ dominated by some $\mathbf{x}_{\text{child}}^i$.

\subsubsection{Backpropagation} 
For each child node $\mathbf{x}_{\text{child}}^i$, we use its reward vector $\mathbf{r}(\mathbf{x}_{\text{child}}^i)$ to initialize the cumulative reward $\mathbf{W}(\mathbf{x}_{\text{child}}^i)=\mathbf{r}(\mathbf{x}_{\text{child}}^i)$ and the number of visits $N_{\text{visit}}(\mathbf{x}_{\text{child}}^i)$ is set to 1. We then trace backward from $\mathbf{x}_{\text{child}}^i$ through its ancestors up to the root node $\mathbf{x}_{\text{root}}$, updating each node $\mathbf{x}$ along the path by accumulating the child rewards and incrementing the visit count:
\begin{align}
    \mathbf{W}(\mathbf{x})&\gets \mathbf{W}(\mathbf{x})+\sum_{i=1}^{M}\mathbf{r}(\mathbf{x}_{\text{child}}^i), \quad N_{\text{visit}}(\mathbf{x})\gets N_{\text{visit}}(\mathbf{x})+1
\end{align}
These cumulative updates influence future selection, prioritizing unmasking paths that lead to higher-reward sequences for further expansion. After $N_{\text{iter}}$ iterations of these four steps, we return the set of Pareto-optimal mRNA sequences optimized across the set of properties $\mathbf{x}^\star\in \mathcal{P}^\star$ and their corresponding score vectors $\mathbf{f}(\mathbf{x}^\star)$. 

\subsection{Plasmid Generation}
\subsubsection{F-Luc and uAb}

The following plasmids were obtained from Addgene: M50 Super 8$\times$ TOPFlash (Addgene \# 12456), M51 Super 8$\times$ FOPFlash (TOPFlash mutant) (Addgene \# 12457), pCMV-3XFLAG-renilla-luciferase-beta-globin-control (Addgene \# 184393), and our pcDNA3-ubiquibody (uAb) cloning vector (Addgene \# 232088). For uAb assembly, oligos for candidate peptides were annealed and ligated via T4 DNA Ligase (NEB, Cat \# M0202S) into the cloning backbones. For plasmid assembly of ABW sequences, equimolar amounts of backbone and insert DNA fragments were combined with 2$\times$ HiFi DNA Assembly Master Mix (NEB, Cat \# E2621) and incubated at 50~$^\circ$C. 

\subsubsection{F-Luc, Spike, and PEMax}

Optimized and benchmark sequences containing 5’ UTR, coding sequences (\textit{P. pyralis} luciferase, SARS-CoV-2 Spike, and PEmax), and 3’ UTR were synthesized as gene fragments and further cloned into a T7 promoter expression vector by GenScript. The final plasmid sequences were confirmed by Sanger sequencing. All mRNAs containing 5’ UTR, ORF, 3’ UTR, and a 103-nucleotide-long poly(A) tail were synthesized via \textit{in vitro} transcription (IVT) by GenScript. 

\subsection{Cell Culture}

HEK293T and DLD-1 cell lines were maintained in Dulbecco’s Modified Eagle’s Medium (DMEM, Gibco, Cat \# 11995073) supplemented with 100 units/mL penicillin, 100\,mg/mL streptomycin (Gibco, Cat \# 15140122), and 10\% fetal bovine serum (FBS, Gibco, Cat \# A5670402) at 37~$^\circ$C in a humidified incubator with 5\% CO$_2$. Jurkat (American Type Culture Collection, TIB-152), A549 (American Type Culture Collection, CCL-185), and HepG2 (American Type Culture Collection, HB-8065) cells were maintained in Roswell Park Memorial Institute 1640 medium (RPMI-1640) (Cytiva, SH30255.01), Dulbecco’s modified Eagle’s medium (DMEM) (Cytiva, SH30243.01), or Eagle's Minimum Essential Medium (EMEM) (ATCC, 30-2003), respectively, and supplemented with 10\% (v/v) fetal bovine serum (Avantor, 97068-085) and 1\% penicillin/streptomycin (Gibco, 15140-122) at 37°C and 5\% CO2. A549 and HepG2 cells were maintained at confluency below 90\%, and Jurkat cells were maintained below a density of $3 \times 10^{6}$ cells per mL. All cells were verified to be Mycoplasma negative (InvivoGen, rep-mys-20).

For luciferase time-course assays, HEK293T cells were seeded at $3.5 \times 10^4$ cells (Figure ~\ref{fig:Fluc}B,C) per well in a white-bottom 96-well plate (Thermo Fisher Scientific, Cat \# 136101). Jurkat, A549, and HepG2 cells were seeded at $1.0 \times 10^4$ cells per well (Figure ~\ref{fig:Fluc}D, E, F) in a 96-well plate. For flow cytometry, A549 cells were seeded at $8.0 \times 10^4$ cells per well (Figure ~\ref{fig:Fluc}G) in a 24-well plate. For \textit{in vitro} PEmax assays, A549 cells were prepared at $2.5 \times 10^5$ per electroporation condition. For quantitative PCR and immunoblotting assays, DLD-1 cells were seeded at $2.5 \times 10^5$ cells per well in 12-well plates 20--24~h before transfection. For TOPFlash-specific samples, DLD-1 cells were seeded at $1.0 \times 10^4$ cells per well in white-bottom 96-well plates (Thermo Fisher Scientific). 100~ng plasmid DNA was added per well in a ratio of TOPFlash (or FOPFlash): Renilla luciferase: pcDNA3 vector = 1:0.1:3; samples were mixed with Lipofectamine\texttrademark~2000 (Invitrogen, Cat \# LMRNA003) in Opti-MEM (Gibco, Cat \# 31985070), as per the manufacturer's recommendations. For all transfection samples, mRNA constructs (100~ng/well for 96-well samples, 500~ng/well for 24-well samples, and 1000~ng/well for 12-well samples) were prepared and transfected with Lipofectamine\texttrademark~MessengerMAX\texttrademark~Transfection Reagent (Invitrogen, Cat \# LMRNA003) in Opti-MEM (Gibco), as per the manufacturer's instructions.  For \textit{in vitro} gene editing assay and PEmax protein expression, A549 cells were electroporated with \textit{in vitro} transcribed PEmax mRNA with or without pegRNA and nicking gRNA using the Neon NxT Electroporation System (Thermo Fisher Scientific) with 10 $\mu$l Neon NxT tips and the corresponding Neon NxT Electroporation Kit (Thermo Fisher Scientific). In brief, $2.5 \times 10^5$ cells were resuspended in 12 $\mu$l of buffer R with 100 ng \textit{in vitro} transcribed PEmax mRNA, 22.5 pmol synthetic pegRNA (GenScript), and 15 pmol synthetic nicking sgRNA (GenScript), or in 12 $\mu$l of buffer R with 1 $\mu$g \textit{in vitro} transcribed PEmax mRNA, for gene editing assay and PE expression, respectively. The cell-RNA mixture was electroporated with the condition of 1600 V, 10 ms, 3 pulses. Following electroporation, cells were suspended in 1 ml of prewarmed complete growth medium in a 24-well plate. For applicable immunoblotting samples, cells were treated with  1~$\mu$M MG132 (proteasome inhibitor, Millipore Sigma, Cat \# M7449) 24~h post-transfection. Transfected or electroporated cells were returned to the incubator at 37°C, 5\% CO2, and processed for downstream assays at the indicated time points. 

\subsection{\textit{In Vitro} Transcription}

Assembled constructs were transformed into 30~$\mu$L NEB Turbo competent \textit{Escherichia coli} cells (NEB, Cat \# C2984H) and plated on LB agar supplemented with the appropriate antibiotic for subsequent sequence verification of colonies and plasmid purification (Genewiz). mRNA encoding Fluc and uAb constructs was synthesized via \textit{in vitro} transcription from PCR-amplified DNA templates containing a T7 promoter using the HiScribe T7 ARCA mRNA Kit (NEB, Cat \# E2060S). The mRNA was then concentrated and cleaned of impurities using the RNeasy MinElute Cleanup Kit (Qiagen, Cat \# 74204) for downstream assays. For plasmids from Genscript, N1-methylpseudouridine (m1$\psi$) (Hongene) instead of UTP was used in the IVT reaction, and the trinucleotide cap1 analogue (Hongene) was capped at the 5’ of mRNA using the co-transcriptional capping method. The IVT products were treated with DNase and then purified with lithium chloride precipitation. All mRNAs were quality-controlled by capillary electrophoresis (TapeStation, Agilent or Fragment Analyzer, Agilent) for integrity and were stored frozen at -20°C. All mRNAs were used within 1 month after being synthesized.

\subsection{Luciferase Time-Course Assay}

Luciferase activity was measured at 12 indicated time points between 0.5 and 60 h post-transfection. Immediately before measurement, IVISbrite D-Luciferin Substrate (Revvity, Cat \# 770505) was added to each well (1:200 dilution), and luminescence signals were collected using a GloMax plate reader (Promega). Luminescence values were background-corrected by subtracting the signal from untreated control wells. Firefly luciferase activity was measured at 7 h, 24 h, 48 h, and 72 h after transfection using the Steady-Glo Luciferase Assay System (Promega) according to the manufacturer’s instructions. Briefly, 100 $\mu$l of reagent was added to cells grown in 100 $\mu$l of medium in each well in the 96-well plate. After incubating at room temperature for 5 minutes, 100 $\mu$l of the mixture was transferred to an opaque white 96-well plate. The luminescence measurements were performed with a SpectraMax iD3 Microplate Reader, measuring the luminescence of all the wavelengths with the endpoint mode. Within each experiment, all samples were measured in technical triplicate.

\subsection{Quantification of SARS-CoV-2 Spike Protein Expression by Flow Cytometry}

Surface expression of SARS-CoV-2 Spike protein in cells was assessed 24 h, 48 h, and 72 h post-transfection by flow cytometry. Cells were harvested, washed once in ice-cold flow staining buffer (DPBS with 0.5\% BSA and 2 mM EDTA), then stained with mouse anti-SARS-CoV-2 Spike (1:200; Cell Signaling, 42172S) for 1 h on ice. Following primary antibody staining, cells were washed twice with flow staining buffer and stained with Anti-mouse IgG (H+L), F(ab')2 Fragment (Alexa Fluor 647 Conjugate) (1:500; Cell Signaling, 4410S) for 30 min on ice. After two final washes with flow staining buffer, surface SARS-CoV-2 Spike levels were analyzed using a BD Accuri C6 Plus Flow Cytometer (BD Biosciences). For each sample, 3,000 live cells were selected based on morphology. SARS-CoV-2 Spike levels were assessed by gating AlexaFluor 647 (excitation at 640nm, collection at 675/25nm) positive cells within the live cells.  Analysis was carried out using BD CSampler Plus. Spike expression level was reported as the mean fluorescence intensity (MFI) of the spike signal among viable cells.

 \subsection{Chemically Synthesized Guide RNAs}
 Single guide RNAs (sgRNAs) and prime editing guide RNAs (pegRNAs) used for \textit{in vitro} gene editing experiments were chemically synthesized by GenScript and were used directly. Both sgRNAs and pegRNAs contained 2'-O-methyl modifications at the first three and last three nucleotides and phosphorothioate linkages between the first three and last three nucleotides. Oligonucleotides were purified by high-performance liquid chromatography (HPLC) and supplied lyophilized.

\subsection{Quantification of Prime Editor Protein Expression by Western Blotting}

PEmax protein expression was determined by Western blot 24 h and 48 h after electroporation of PEmax mRNA. Cells were washed with ice-cold DPBS and lysed in RIPA buffer (Thermo Fisher Scientific) supplemented with Halt 100X protease inhibitor (Thermo Fisher Scientific). Protein concentrations were determined using the Pierce BCA Protein Assay Kit (Thermo Fisher Scientific). Equal amounts of total protein (10 µg per lane) were denatured and separated by SDS-PAGE (Thermo Fisher Scientific) and transferred to 0.22 µm polyvinylidene difluoride (PVDF) membranes using the eBlot L1 Transfer system (GenScript). Membranes were blocked for 1 h at room temperature in 5\% non-fat milk in TBST, cut at around 100kDa based on the protein ruler, then incubated with mouse anti-Cas9 (1:1,000; Cell Signaling, 14697S) or mouse anti-beta actin (1:10,000; Proteintech, 66009-1-Ig) overnight at 4 °C. After three washes in TBST, membranes were incubated for 1 h at room temperature with HRP-conjugated secondary antibody (1:10,000; Thermo Fisher Scientific, 31430), followed by another three washes in TBST. Pierce ECL Western Blotting Substrate (Thermo Fisher Scientific) was added onto the membranes, and the membranes were imaged on Syngene G:BOX Imager to detect chemiluminescent signals. Band intensities were quantified using ImageJ, and the PEmax signal was normalized to the beta actin signal within the same lane, and then to the benchmark (non-optimized sequence) on the same blot. 

\subsection{Quantification of Editing Efficiency by Sanger Sequencing}
Editing efficiency at the HEK site 3 locus was quantified from bulk cell populations harvested 72 h after electroporation. Genomic DNA was extracted from the cells using QuickExtract DNA Extraction Solution (Biosearch Technologies) following the manufacturer’s instructions. Locus-specific primers (forward primer: CAGGGAAACGCCCATGCAATTAG; reverse primer: CTCTGTTGAGCTCGACCCTGAA) were used to generate targeted amplicons for Sanger sequencing. The target region was amplified from the genomic DNA in a 20 µl reaction using OneTaq 2X Master Mix (New England Biolabs, NEB, M0482S) with the following conditions: 94 °C for 2 min; 27 cycles of 94°C for 20s, 60°C for 20s, and 68°C for 20s, followed by 68°C for 1 min. PCR products were purified using Hieff NGS DNA Selection Beads (Yeasen) following the manufacturer’s instructions. Purified amplicons were submitted to Eurofins Genomics for Sanger sequencing using the reverse PCR primer. The editing efficiency at the intended edit position was quantified using the EditR (https://moriaritylab.shinyapps.io/editr\_v10/) with default parameters.

\subsection{Quantitative PCR}

 Cells were harvested at 9 h and 24 h post-transfection, and total RNA was isolated using the Monarch\textsuperscript{\textregistered} Spin RNA Isolation Kit (Mini) (NEB, Cat \# T2110S). RNA concentration was determined by NanoDrop and normalized before reverse transcription. cDNA was synthesized from normalized RNA using LunaScript\textsuperscript{\textregistered} RT SuperMix Kit (NEB, Cat \# E3010L), and the resulting cDNA was diluted before qPCR analysis. Quantitative PCR was performed using PowerUp\texttrademark~SYBR\texttrademark~Green Master Mix (Thermo Fisher Scientific, Cat \# A25776) in technical triplicates with gene-specific primers. Relative transcript levels were normalized to the indicated housekeeping gene (GAPDH) and analyzed using the $\Delta\Delta Ct$ method. Primers for each target gene were as follows: GAPDH: forward: ATGGGGAAGGTGAAGGTCG, reverse: TCCCGTTCTCAGCCTTGACG, $\beta$-catenin uAb: forward: GATCTGTACATCCCGGCCTTC, reverse: GCCTGGAGAGGTAGGAGTGTA.

\subsection{TOPFlash Assay}

48 h post-transfection, cells were lysed, and firefly and Renilla luciferase activities were measured sequentially using the Dual-Luciferase Reporter Assay System (Promega, Cat \# E1980). Firefly luciferase activity was normalized to the corresponding Renilla signal, and all values were subsequently normalized to the TOPFlash + polyG uAb control condition.

\subsection{Cell Fractionation and Immunoblotting}

72 h post-transfection, DLD-1 cells were detached with 0.05\% trypsin-EDTA, washed twice with ice-cold 1$\times$ PBS, and lysed in Pierce\texttrademark~RIPA Buffer (Thermo Fisher Scientific, Cat \# 89900) supplemented with a 1:100 dilution of protease inhibitor cocktail (Millipore Sigma, Cat \# P8340). Specifically, the protease inhibitor cocktail--RIPA buffer solution was added to the cell pellet, and the mixture was incubated at 4~$^\circ$C for 30 min, followed by centrifugation at 12{,}000~rpm for 20 min at 4$^\circ$C. The supernatant was collected into tubes, and protein concentrations were quantified using the Pierce\texttrademark~BCA Protein Assay Kit (Thermo Fisher Scientific, Cat \# 23227). 25 mg of total protein were mixed with 4$\times$ Bolt\texttrademark~LDS Sample Buffer (Thermo Fisher Scientific, Cat \# NP0007) containing 5\% $\beta$-mercaptoethanol (Millipore Sigma, Cat \# M3148) at a 3:1 ratio and heated at 95~$^\circ$C for 10~min prior to electrophoresis. Samples were loaded at equal volumes into Bolt\texttrademark~Bis-Tris Plus Mini Protein Gels (Thermo Fisher Scientific, Cat \# NW04125BOX) and separated by electrophoresis. Immunoblotting was performed according to standard protocols. iBlot\texttrademark~2 Dry Blotting System with iBlot\texttrademark~2 Transfer Stacks (Invitrogen) were used for membrane blot transfer. Membranes were then blocked in 5\% milk in 1$\times$ TBST for 1 h at room temperature and incubated overnight at 4~$^\circ$C with rabbit anti-$\beta$-catenin antibody (Cell Signaling Technology, Cat \# 8480S; diluted 1:1000) or mouse anti-GAPDH antibody (Santa Cruz, Cat \# sc-47724; diluted 1:2500) for loading control. After three washes with 1$\times$ TBST for 5~min each, membranes were incubated with a secondary antibody, goat anti-Mouse IgG (H+L), Alexa Fluor\texttrademark~488 (Thermo Fisher Scientific, Cat \# A11001; diluted 1:5000) or donkey anti-Rabbit IgG, IRDye 800 (Fisher Scientific, Cat \# NC9523609; diluted 1:4000), for 1~h at room temperature. Following three additional washes with 1$\times$ TBST, blots were detected by fluorescence signals using a LICOR Odyssey\textsuperscript{\textregistered} Imager. Densitometry analysis in immunoblots was performed using ImageJ software. Briefly, bands in each lane were grouped as a horizontal lane and quantified using FIJI’s gel analysis function. $\beta$-catenin band intensities were normalized to the band intensity of GAPDH in each lane, then to the average band intensity for untreated control cases across replicates.

\subsection{Statistical Analysis and Reproducibility}

Sample sizes were not predetermined based on statistical methods but were chosen according to standards of the field (three independent biological replicates for each condition, where possible), which provided sufficient statistical power for the effect sizes of interest. All data are reported as mean values with error bars representing standard deviation (SD). Statistical analyses were performed using the two-tailed Student's $t$-test in GraphPad Prism 10 software, with calculated $p$ values represented as follows: *$p \leq 0.05$, **$p \leq 0.01$, ***$p \leq 0.001$, ****$p \leq 0.0001$. No data were excluded from the analyses except for one technical outlier that was removed based on predefined criteria. The experiments were not randomized, and the investigators were not blinded to allocation during experiments or outcome assessment.

\section*{Declarations}

\paragraph{Data and Code Availability}
mRNAutilus is hosted and available as a part of AutoNA (Automatic Nucleic Acid) at \href{https://autona.atombio.ai}{https://autona.atombio.ai}. Pretraining and regressor data is deposited at \href{https://doi.org/10.5281/zenodo.19600209}{10.5281/zenodo.19600209}.

\paragraph{Acknowledgments} We thank the entire experimental team of the Chatterjee Lab and Atom Bioworks for productive input during the curation of these models and datasets. We also thank Lauren Hong for designing the mRNAutilus logo. 

\paragraph{Author Contributions} S.P. and S.T. developed the mRNAutilus algorithm and framework and performed all \textit{in silico} benchmarking, with assistance from Y.Z. Y.K. and P.J.L. led experimental validation, with assistance from D.S, S.S, and F.P. F.P., C.W., S.Y., and P.C. supervised the work.  S.P., S.T., Y.K., and P.C. wrote the manuscript, with input from all authors. S.P. proposed the initial idea.

\paragraph{Funding Statement} This research was supported by a grant from the Hartwell Foundation to the lab of P.C., as well as funding from Atom Bioworks, Inc. We also acknowledge a seed grant from the ICLR Workshop on Generative and Experimental Perspectives for Biomolecular Design to the lab of P.C. and Atom Bioworks, Inc.

\paragraph{Competing Interests} S.Y. and P.C. are co-founders of Atom Bioworks, Inc., which is involved in mRNA development. S.P. is an employee of Atom Bioworks, Inc. P.C. is a co-founder of Gameto, Inc., UbiquiTx, Inc. P.C.’s interests are reviewed and managed by the University of Pennsylvania in accordance with their conflict-of-interest policies. P.J.L, S.S., F.P., and C.W. are employees of GenScript, Inc. The remaining authors have no conflicts of interest to declare.

\bibliographystyle{unsrtnat}
\bibliography{citation}  

\newpage

\appendix
\renewcommand{\thefigure}{S\arabic{figure}}
\setcounter{figure}{0}
\setcounter{table}{0}
\renewcommand{\thetable}{S\arabic{table}}

\newpage
\newpage
\begin{center}
    {\LARGE\bfseries\color{PennBlue} Supplementary Information}
\end{center}
\vspace{1em}

Supplement \ref{appendix:A} presents additional results, including unconditional generation metric distributions for CDSs (~\ref{appendix:A.1}) and UTRs (~\ref{appendix:A.2}), an investigation of the mRNA pretraining dataset, principal component analysis of the latent embeddings learned by our unconditional MDM (~\ref{appendix:A.4}), the hyperparameters for the XGBoost regressors (~\ref{appendix:A.5}), XGBoost regressor comparison against KNN (~\ref{appendix:A.6}), the guidance curves for human MUC1 and SARS-CoV-2-S-Protein mRNA generation (\ref{appendix:A.7}), comparative evaluation for property prediction against other language models (~\ref{appendix:A.8}), and the protocol for producing the full-length mRNA sequences tested \textit{in vitro} (~\ref{appendix:A.9}). We also provide pseudo code for our multi-objective guidance algorithm in Appendix \ref{appendix:B}.

\section{Additional Results}
\label{appendix:A}
\subsection{Unconditional Generation}
\label{appendix:A.1}
Before applying multi-objective guidance, we evaluated the ability of mRNAutilus to unconditionally generate mRNAs that resemble natural mRNAs by comparing them to natural eukaryotic mRNAs. Here, we generated three libraries of 200 mRNAs each, with token lengths of either 500b, 1000b, or 1,500b. These generated libraries were compared with natural eukaryotic mRNAs from the Ensembl pretraining dataset of near-equal length. Equal length was not guaranteed, as it was possible for the model to asymmetrically sample codon tokens over UTR tokens, resulting in inflated length. As such, we primarily consider length-normalized metrics such as GC Content and length-normalized minimum free energy. 

\begin{figure}[t]
    \centering
    \includegraphics[width=\linewidth]{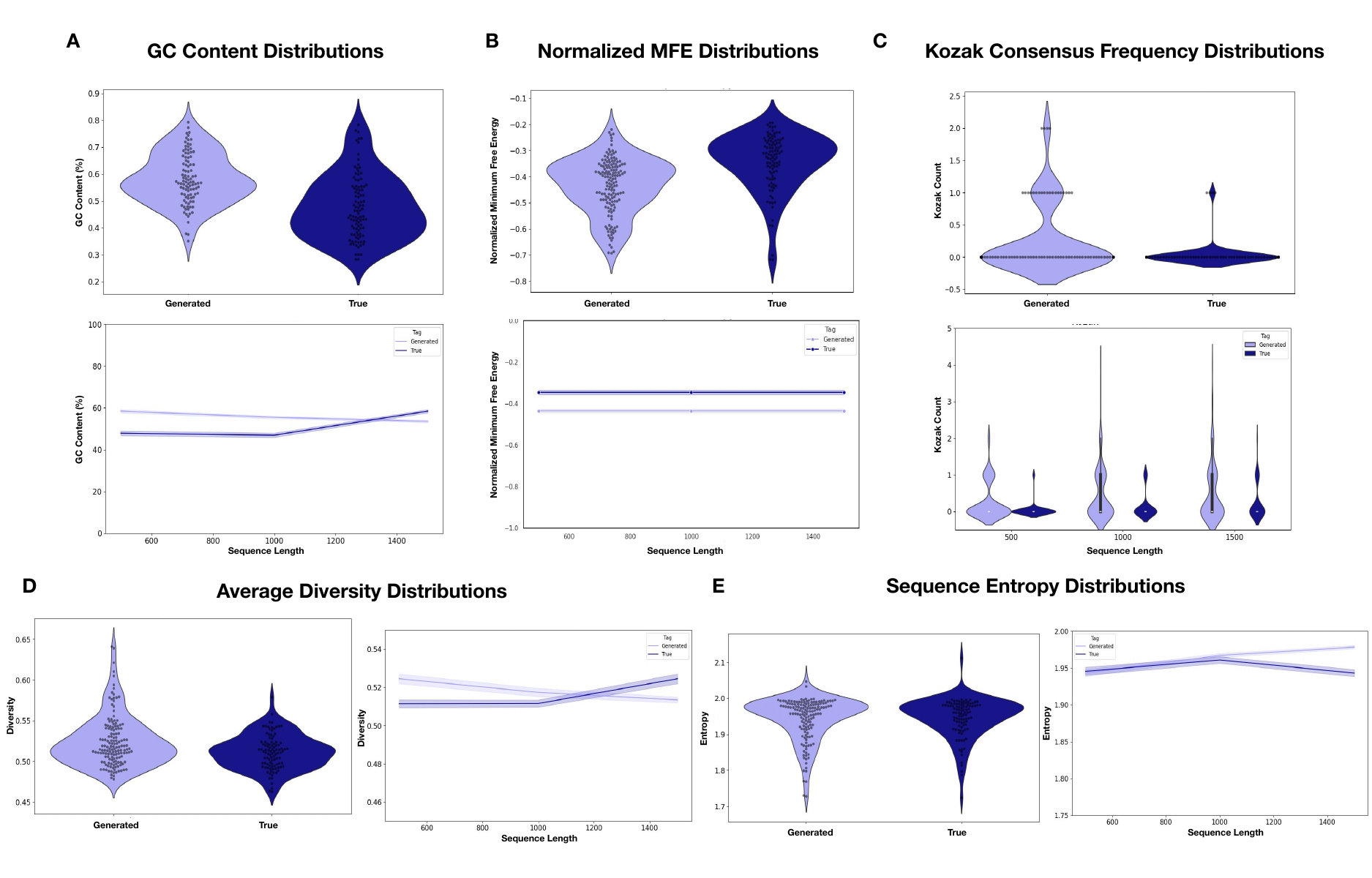}
    \vspace{-10pt}
    \caption{\textbf{Unconditional mRNA generation results.} \textbf{(A)} GC content, \textbf{(B)} minimum free energy normalized by sequence length, and \textbf{(C)} Kozak motif frequency evaluations on unconditionally-generated sequences. \textbf{(D)} Average sequence diversity and \textbf{(E)} Shannon entropy evaluations on unconditionally-generated sequences. Sequences of length 1,000 have their distributions shown (left), while the average values for each property at each length are shown as line plots (right). The error bars on the line plots denote the standard error.}
    \label{fig:unconditional}
\end{figure}
\subsection{Generated UTR evaluation}
\label{appendix:A.2}

\begin{figure}[ht]
    \centering
    \includegraphics[width=\linewidth]{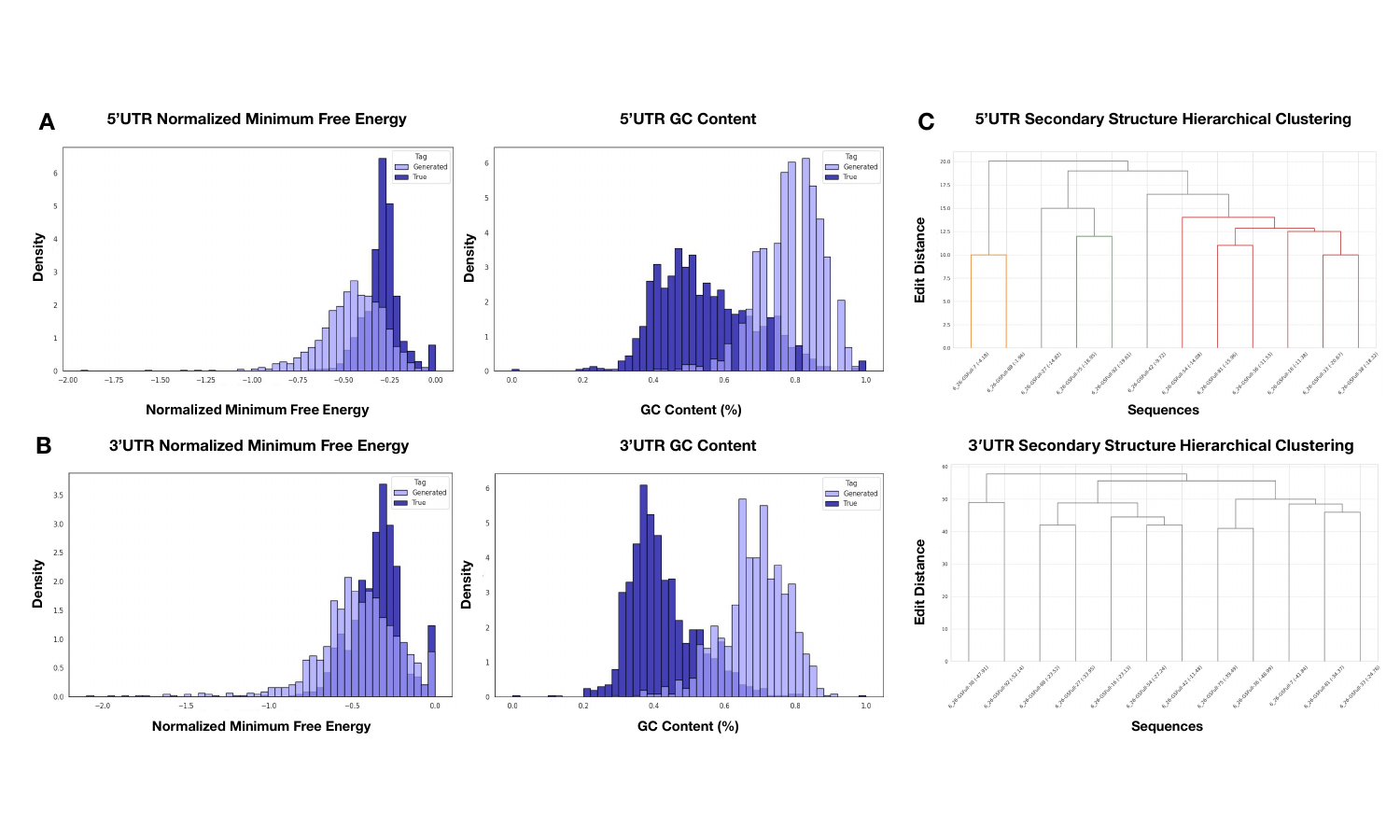}
    \vspace{-10pt}
    \caption{\textbf{Analysis of 5’ and 3’UTRs for \textit{P. pyralis} luciferase mRNAs.} \textbf{(A)} Normalized minimum free energy (left) and GC content (right) distributions (left) for 1,000 generated 5’UTRs in comparison with 1,000 vertebrate 5’UTRs from Ensembl. \textbf{(B)} Normalized minimum free energy (left) and GC content (right) distributions (left) for 1,000 generated 3’UTRs in comparison with 1,000 vertebrate 3’UTRs from Ensembl. \textbf{(C)} Hierarchical clustering of predicted 5’ (top) and 3’ (bottom) secondary structures for 12 generated \textit{P. pyralis} luciferase mRNAs.}
    \label{fig:utrs}
\end{figure}

We also further characterized the UTRs designed by mRNAutilus in comparison to those designed through evolution in Figure ~\ref{fig:utrs}. Notably, we observe that the generated 5'UTRs have slightly more positive predicted MFEs and GC contents than those existing in nature. More positive MFE values in RNA indicate reduced secondary structure, which has been associated with improved translation efficiency \citep{mauger2019mrna}.

\subsection{Pretraining dataset}
\label{appendix:A.3}
\begin{figure}[h]
    \centering
    \includegraphics[width=\linewidth]{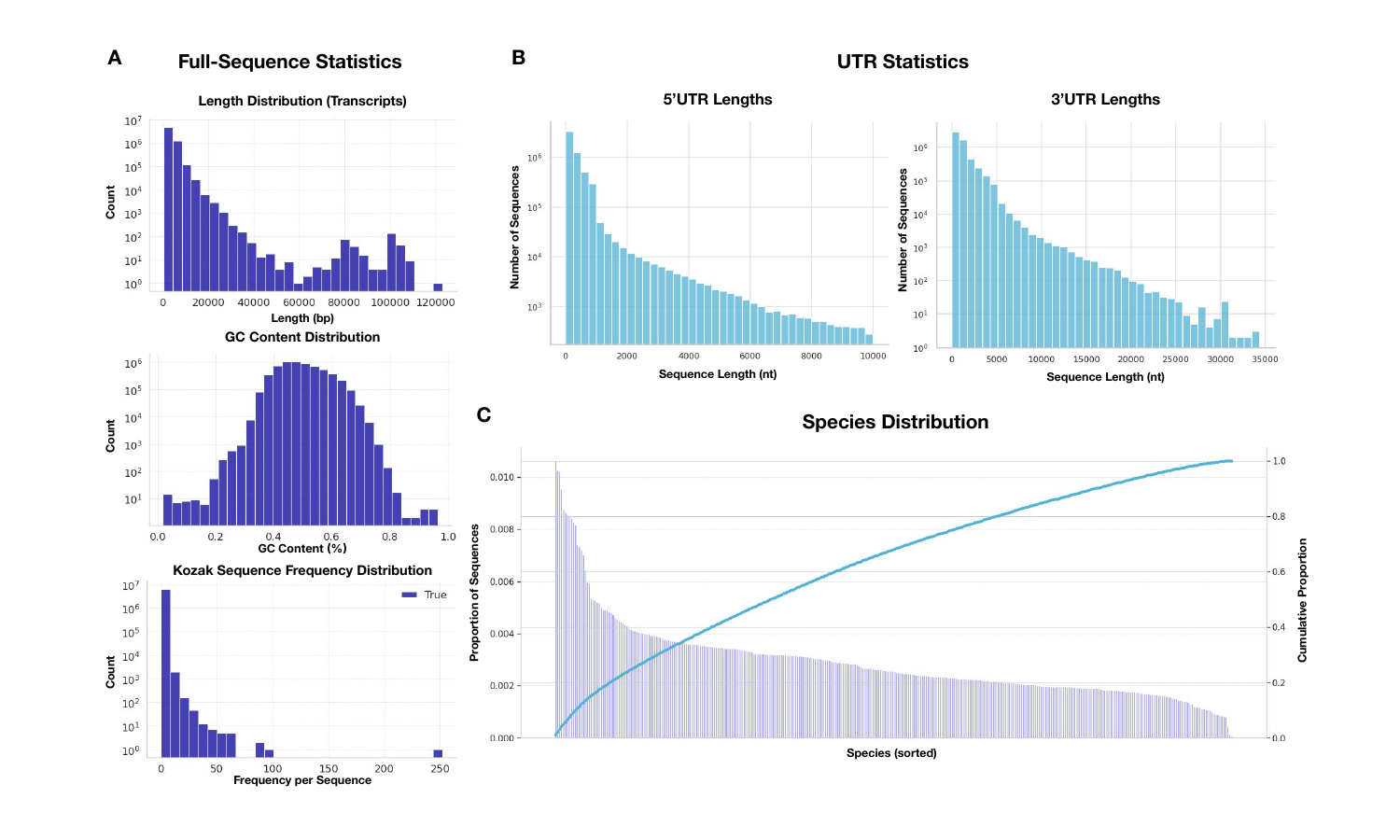}
    \vspace{-10pt}
    \caption{\textbf{mRNA pretraining data.} \textbf{(A)} Distributions for all ~5.5 million protein-coding transcripts in the Ensembl dataset for sequence length (top-left), GC content (middle-left), and Kozak consensus sequence frequency (bottom-left). \textbf{(B)} Sequence length distributions for the 5' (left) and 3'UTRs (right) of all sequences used during pretraining. \textbf{(C)} The proportion of protein-coding transcripts attributed to each of the 342 vertebrate species in the pretraining corpus.}
    \label{fig:pretraining}
\end{figure}

In Figure ~\ref{fig:pretraining}, we investigated metrics distributions for the component CDSs and UTRs for all vertebrate mRNAs included in the Ensembl pretraining corpus. We consider sequence-level quantities such as length, GC-content, and frequency of Kozak motifs per-sequence as defined by the regular expression \texttt{(A|G)CCATGG}. We additionally consider the lengths of both UTR populations and the distribution of mRNAs across all mammalian species considered in the Ensembl database.

\subsection{PCA Analysis}
\label{appendix:A.4}

\begin{figure}[t]
    \centering
    \includegraphics[width=\linewidth]{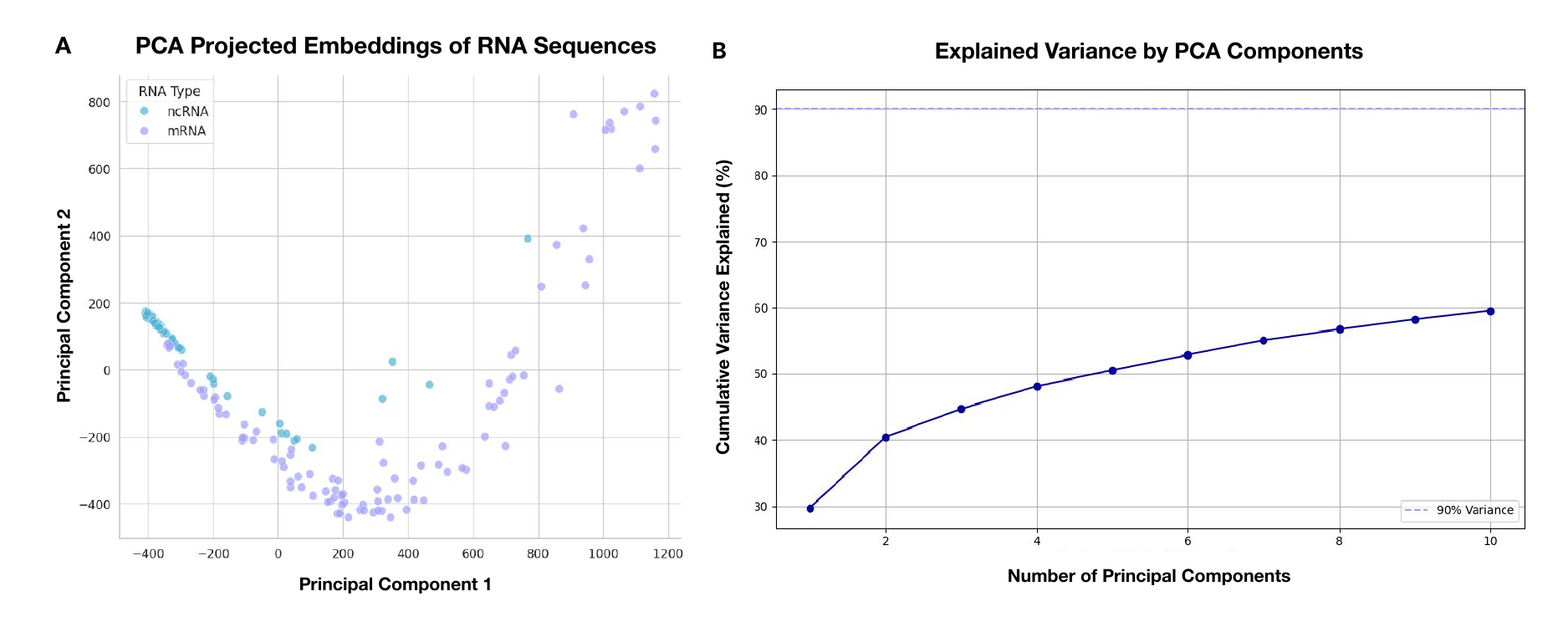}
    \vspace{-10pt}
    \caption{\textbf{PCA Analysis of mRNAutilus representations.} \textbf{(A)} mRNAutilus embeddings are collected for 100 mRNAs and ncRNAs each, projected onto the two-dimensional vector space defined by the first two principal components of the flattened embeddings. \textbf{(B)}The associated scree plot for the PCA.}
    \label{fig:pca}
\end{figure}

To investigate the representations provided by mRNAutilus, we compared the model-produced embeddings for 100 mRNA sequences and 100 non-coding RNA (ncRNA) sequences. The mRNAs were sourced from the pretraining data validation set, and the non-coding RNAs were taken from RNACentral \citep{rnacentral2021rnacentral}. All 200 sequences were embedded and flattened, after which they were projected onto the first two principal components of the resulting two-dimensional matrix. Per Figure ~\ref{fig:pca}, the mRNAs are separable from the ncRNAs.

\subsection{Regressor Hyperparameters}
\label{appendix:A.5}
Hyperparameters for the XGBoost regressors are included in Table ~\ref{tab:xgboost_params_regression}. All regressors were trained on the data described in the Results section ~\ref{data-collection}, without any modifications to the sequences or pre-processing. Regressor training/validation sets were 80/20\% splits done at random.
\begin{table}[h!]
\centering
\caption{XGBoost Hyperparameters for Regression}
\label{tab:xgboost_params_regression}
\vskip 0.05in
\begin{tabular}{@{}ll@{}}
\toprule
\textbf{Hyperparameter} & \textbf{Value/Range} \\ \midrule
Objective & \texttt{reg:squarederror} \\
Lambda  & \([0.1, 10.0]\) (log scale) \\
Alpha & \([0.1, 10.0]\) (log scale) \\
Gamma & \([0, 5]\) \\
Colsample by Tree & \([0.5, 1.0]\) \\
Subsample & \([0.6, 0.9]\) \\
Learning Rate & \([1\text{e}{-5}, 0.1]\) \\
Max Depth & \([2, 30]\) \\
Min Child Weight & \([1, 20]\) \\
Tree Method & \texttt{hist} \\
Scale Pos Weight  & \([0.5, 10.0]\) (log scale) \\
\bottomrule
\end{tabular}
\end{table}

\subsection{Property Prediction}
\label{appendix:A.6}
To facilitate property prediction using mRNAutilus language model embeddings, we considered three regressor implementations: linear probing, XGBoost, and K-Nearest Neighbors (KNN). We demonstrate the superiority of the XGBoost-based implementation in Table ~\ref{table:knn}. Both methods were implemented using scikit-learn and Optuna. 
\begin{table}
\caption{\textbf{Comparison of different mRNA property prediction methods.} Each method was implemented using Optuna to fit an optimized regression model according to the constituent hyperparameters. In both implementations, sequences were truncated to 5,500 tokens before being embedded by mRNAutilus and used to fit either model.}
\label{table:knn}
\begin{center}
\resizebox{0.8\linewidth}{!}{
\begin{tabular}{@{}llccccccc@{}}
\toprule
\textbf{Regressor} &\multicolumn{2}{c}{\textbf{Half-Life}} & \multicolumn{2}{c}{\textbf{Translation Efficiency (TE)}} & \multicolumn{2}{c}{\textbf{Protein Abundance}} \\ \midrule
 & Train/$R^2$ & Val/$R^2$ & Train/$R^2$ & Val/$R^2$ & Train/$R^2$ & Val/$R^2$ \\
\midrule
XGBoost  & $0.949$ & $0.551$ & $0.923$ & $0.716$ & $0.926$ & $0.565$ \\
KNN & $1.0$ & $0.486$ & $1.0$ & $579$ & $1.0$ & $0.511$ \\
\bottomrule
\end{tabular}
}
\end{center}
\end{table}

\subsection{Conditional guidance}
\label{appendix:A.7}
\paragraph{\textit{P. pyralis} luciferase protein}
\textit{P. pyralis} luciferase, or Fluc, is an enzyme that catalyzes the oxidation of firefly luciferin, yielding bioluminescence. Due to this natural luminescence mechanism, it is commonly used in research as a reporter for evaluating the transcriptional activity profile in cells transfected with a construct containing the luciferase gene controlled by a promoter of interest. As a commonly used protein in R\&D efforts, we designed a more optimal \textit{P. pyralis} luciferase mRNA using mRNAutilus. The template luciferase was constructed referencing the GenScript F-Luc mRNA ORF \citep{trilink_fluc_mrna}. The lengths of the UTRs were arbitrarily set at 56 and 114 nt, respectively. 

We performed guided design of Fluc mRNA by masking the coding sequence at a 15\% masking rate uniformly at random and fully masking the UTRs. The Pareto front size was set to 24 sequences, with the number of MCTS iterations set to 200.

\paragraph{SARS-CoV-2 S-Protein}

The S-Protein is a trimeric class I fusion protein on the surface of the SARS-CoV-2 virus, which directly mediates viral entry into host cells. It contains several subunits, notably the receptor-binding domain (RBD), which binds the human ACE2 receptor. This protein, albeit across different isoforms, is the primary target for all current therapeutic efforts against COVID-19. For the design task, we constructed a template mRNA transcript for the wild-type S-protein around the publicly available CDS \citep{NCBI_NC_045512.2}. Referencing an S-Protein mRNA vaccine patent \citep{WO2021159040A2}, we set lengths 56 and 114 for the 5' and 3'UTRs, respectively.

We performed simultaneous 5'/3' UTR design and partial codon optimization of the CDS, with uniform masking at a 15\% mask rate over the entire frame. The Pareto front size was set to 12 sequences, with the number of MCTS iterations set to 128.

\paragraph{Human MUC1 transmembrane protein}
The MUC1 transmembrane glycoprotein is typically expressed on the apical surface of epithelial cells and is responsible for lubrication, protection, and cell signaling. In epithelial cancers, such as in the lung, pancreas, ovaries, breast, and colon, MUC1 is overexpressed and atypically glycosylated. Its characteristic profile in the context of cancers qualifies it as a tumor-associated antigen. We define a template with the entire wild-type CDS, which we partially optimize, and set lengths to 66 and 311 for the 5' and 3'UTRs, respectively. We performed simultaneous 5'/3' UTR design and partial codon optimization of the CDS, with uniform masking at a mask rate of 15\% throughout the frame. The Pareto front size was set to 12 sequences, with the number of MCTS iterations set to 128.

\begin{figure}[t]
    \centering
    \includegraphics[width=\linewidth]{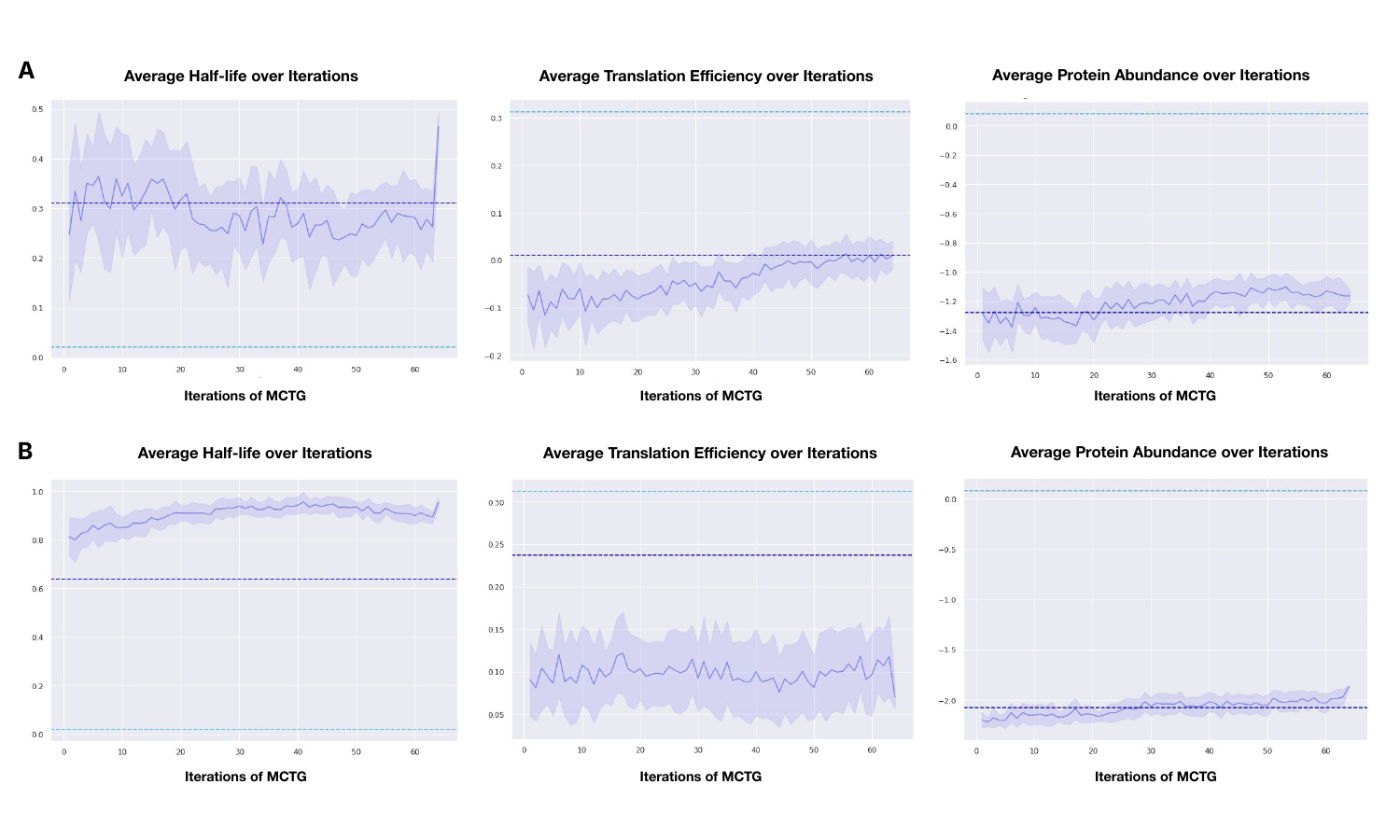}
    \vspace{-10pt}
    \caption{\textbf{Conditional generation results for (A) Human MUC1 and (B) SARS-CoV-2 S-Protein.} Conditional sequence generation spanned 64 iterations, with a Pareto front size set to 12. Navy and teal lines correspond to the regressor dataset medians and the human alpha-globin (HAB)-UTR mRNA regressor scores, respectively. Error bars denote the standard error across all sequences in the Pareto front.}
    \label{fig:cond-extra}
\end{figure}

For the SARS-CoV-2 S-Protein and human MUC1 mRNA generation, we produced 12 Pareto-optimal mRNAs using 3 property-centric regressors over 128 MCTS iterations. Results for conditional generation are shown in Figure ~\ref{fig:cond-extra}.

\begin{table}[t]
\caption{\textbf{Comparison of property scores for wild-type mRNA transcript and codon optimized mRNA transcript with fully-designed UTR sequences.} 24 sequences were generated for the Fluc template, and 12 were generated for the S-Protein and MUC1 templates. The scores of a single optimized transcript with the highest overall scores are shown in comparison to its wild-type counterpart.}
\label{table:Results}
\begin{center}
\resizebox{\linewidth}{!} {
\begin{tabular}{@{}llccccccc@{}}
\toprule
\textbf{Template Gene} &\multicolumn{2}{c}{\textbf{Half-Life (log-10 hours; $\uparrow$)}} & \multicolumn{2}{c}{\textbf{Translation Efficiency (TE) ($\uparrow$)}} & \multicolumn{2}{c}{\textbf{Protein Abundance (log-10 scale; $\uparrow$)}} \\ \midrule
 & wild-type & designed & wild-type & designed & wild-type & designed \\
\midrule
Fluc  & $-0.112$ & $\mathbf{0.537}$ & $-0.034$ & $\mathbf{0.283}$ & $-1.22$ & $\mathbf{-0.451}$ \\
SARS-CoV-2-S-Protein & $-0.590$ & $\mathbf{1.074}$ & $0.006$ & $\mathbf{0.236}$ & $-0.741$ & $\mathbf{-0.727}$ \\
MUC1  & $-0.092$ & $\mathbf{0.593}$ & $0.089$ & $\mathbf{0.092}$ & $\mathbf{-0.630}$ & $-0.906$ \\
\bottomrule
\end{tabular}
}
\end{center}
\end{table}

\begin{figure}[h!]
    \centering
    \includegraphics[width=\linewidth]{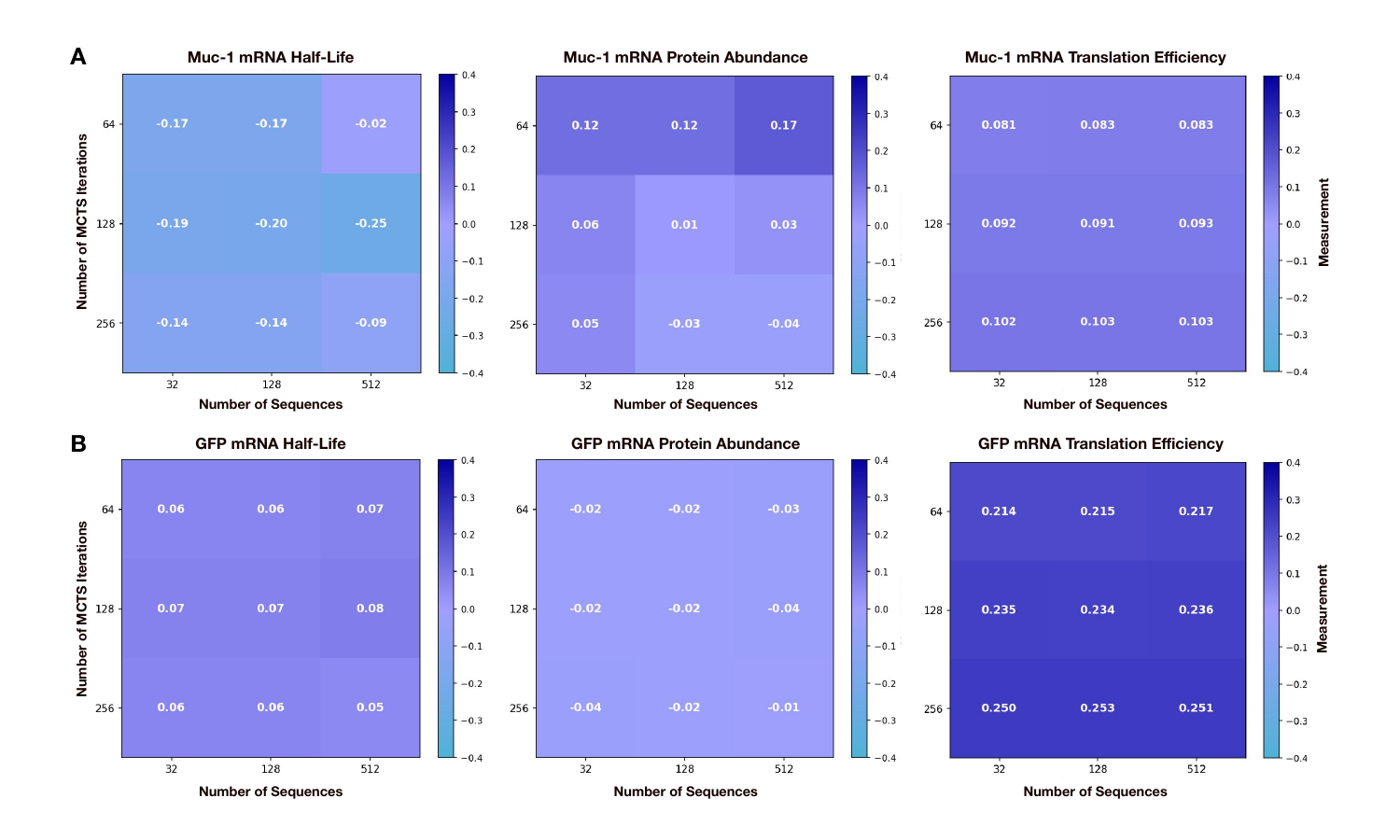}
    \vspace{-10pt}
    \caption{\textbf{Ablation of the key MCTG parameters.} A gridded conditional generation was performed using an increasingly larger number of MCTG iterations and the number of sequences in the Pareto front. We demonstrate this study for \textbf{(A)} MUC-1, a shorter mRNA with a CDS of length 795, and \textbf{(B)} GFP, a longer mRNA with a CDS of length 3,821.
}
    \label{fig:ablation}
\end{figure}

We also perform an ablation on key parameters for the MCTG procedure, including the number of MCTG iterations and the number of sequences in the Pareto front. We demonstrate results for two target genes of different lengths - MUC-1 and GFP. Results are shown in Figure ~\ref{fig:ablation}, where we find that a reduced number of MCTG iterations tends to correlate with greater mRNA fitness in the Pareto front.

\subsection{Comparative Evaluation}
\label{appendix:A.8}
\begin{table}[ht]
\centering
\caption{\textbf{Benchmarking property prediction across models.} We report the Validation $R^2$ for each predictive task, where annotated half-life, translation efficiency, and protein abundance datasets are embedded using each language model and used to train an XGBoost regressor for recapitulating the experimentally-determined value.}
\label{tab:benchmarking}
\vspace{0.5em}
\resizebox{\linewidth}{!}{
\begin{tabular}{lcccc}
\toprule
\textbf{Model} & \textbf{Parameters} & \textbf{Half-Life ($R^2$, ↑)} & \textbf{Translation Efficiency ($R^2$, ↑)} & \textbf{Protein Abundance ($R^2$, ↑)} \\
\midrule
Helix-mRNA     & $6$M & $0.481$ & $0.767$ & $0.544$ \\
Hyena-DNA      & $7$M & $0.426$ & $0.514$ & $0.518$ \\
RiNALMo        & $150$M & $0.497$ & $0.682$ & $0.527$ \\
mRNAutilus     & $150$M & $\mathbf{0.551}$ & $0.716$ & $0.565$ \\
Evo-2          & $7$B & $0.538$ & $\mathbf{0.794}$ & $\mathbf{0.566}$ \\
\bottomrule
\end{tabular}
}
\vspace{-0.5em}
\end{table}

Following the same methodology above for property prediction, we additionally trained XGBoost regressors using other commonly-used nucleic acid language model embeddings, namely Hyena-DNA-large-1M \citep{nguyen2023hyenadna}, RiNALMo-150M \citep{penic2025rinalmo}, and Evo-2 \citep{brixi2026genome}. We additionally include Helix-mRNA \citep{wood2025helix} as the only open-sourced pretrained mRNA language model, to our knowledge. The aforementioned models were also trained on full-length nucleic acid sequences, similar to mRNAutilus. They also tokenize nucleic acid sequences at single-nucleotide resolution, which was maintained in our evaluation. All sequences were kept to a maximum length of 5,500 tokens in our benchmarking. Results on the validation set are shown in Table ~\ref{tab:benchmarking}.

\subsection{\textit{In vitro}-tested mRNA sequences}
\label{appendix:A.9}

\begin{table}[t]
\centering
\caption{\textbf{Generated sequences optimized for half-life and translation efficiency in human cell lines achieve superior property prediction distributions over the baseline human alpha-globin UTR construct.} 500 mRNAs per gene were produced via the MCTG sampling algorithm and scored according to XGBoost property regressors predicting mRNA half-life and translation efficiency.}
\label{table:invitro}
\resizebox{0.8\linewidth}{!} {
\begin{tabular}{@{}llccccccc@{}}
\toprule
\textbf{Template Gene} &\multicolumn{2}{c}{\textbf{Half-Life (log-10 hours; $\uparrow$)}} & \multicolumn{2}{c}{\textbf{Translation Efficiency (TE) ($\uparrow$)}} \\ \midrule
 & Baseline & Designed & Baseline & Designed \\
\midrule
Fluc  & $0.68142$ & $\mathbf{0.95149 \pm 0.05839}$ & 0.10435 & $\mathbf{0.28258 \pm 0.05600}$\\
SARS-CoV-2-S-Protein & $0.41234$ & $\mathbf{0.67300 \pm 0.05917}$ & $0.05752$ & $\mathbf{0.26410 \pm 0.02014}$ \\
PEMax & $\mathbf{0.59335}$ & $0.58568 \pm 0.03314$ & $0.02213$ & $\mathbf{0.02613 \pm 0.01122}$ \\
uAb & $0.06028$ & $\mathbf{0.46191 \pm 0.02740}$ & $-0.55061$ & $\mathbf{0.02371 \pm 0.00896}$ \\
\bottomrule
\end{tabular}
}
\end{table}

To further evaluate our generated mRNAs, we generated libraries (N=200) encoding \textit{P. pyralis} luciferase, SARS-CoV-2 Spike glycoprotein, PEMax gene editing enzyme, and $\beta$-catenin uAb. For all, we used the translation efficiency and half-life regressors for reward computation. Additionally, sequences were produced following 64 iterations during the MCTG algorithm and sampled following 64 diffusion timesteps. Codon optimization was considered for 20\% of codons in the CDS, and all UTRs were designed simultaneously. Template ORFs were taken from GenScript F-Luc mRNA, GenScript Spike V2, GenScript PEMax, the $\beta$-catenin uAb mRNA PpC$_3$ reported in \citet{bhat2025novo}. In total, we produced 500 mRNAs for each gene and synthesized the top 4 or 5 sequences according to overall fold improvement in comparison to the HAB baseline. Score distributions for each gene and for each property are shown in Table ~\ref{table:invitro}.

\clearpage
\section{Algorithm}
\label{appendix:B}
Here, we provide the pseudocode for Monte-Carlo Tree Guidance (MCTG) used in the mRNAutilus framework.
\begin{algorithm}[h!]
\small
\caption{mRNAutilus with Monte-Carlo Tree Guidance}
\label{alg:MCTS}
\begin{algorithmic}[1]
    \State \textbf{Input:} Denoising model $p^\theta_{0|t}(\mathbf{x}_0|\mathbf{x}_t)$, score function $\mathbf{f}(\mathbf{x}) : \mathcal{V}^L \to \mathbb{R}^K$, number of iterations $N_{\text{iter}}$, number of children $M$
    \State \textbf{Output:} Set of Pareto-optimal mRNA sequences $\mathcal{P}^\star$
    \State \textbf{Initialize:} $\mathbf{x}_1 \gets [\mathbf{m}]^L$, $\mathcal{P}^\star \gets \{\}$, $t \gets 1$
    
    \For{$i = 1$ to $N_{\text{iter}}$}
        \State \colorbox{gray!20}{\textbf{Selection:}}
        \LComment{Traverse tree to expandable leaf}
        \State $\mathbf{x}_{\text{leaf}} \gets \textsc{Select}(x_0)$
        
        \State \colorbox{gray!20}{\textbf{Expansion:}}
        \For{$j = 1$ to $M$}
            \State Sample $\mathbf{x}_0^{(j)} \sim p^\theta_{0|t}(\cdot | \mathbf{x}_{\text{leaf}})$ with Gumbel noise
            \State Create child $\mathbf{x}_{\text{child}}^{(j)}$ by unmasking $k$ positions according to $\mathbf{x}_0^{(j)}$
            \State Add $\mathbf{x}_{\text{child}}^{(j)}$ to $\text{children}(\mathbf{x}_{\text{leaf}})$
        \EndFor
        
        \State \colorbox{gray!20}{\textbf{Rollout:}}
        \For{each $\mathbf{x}_{\text{child}}^{(j)} \in \text{children}(\mathbf{x}_{\text{leaf}})$}
            \LComment{Complete sequence using self-planned ancestral sampling}
            \State $\tilde{\mathbf{x}}^{(j)} \gets \texttt{SelfPlannedSampling}(\mathbf{x}_{\text{child}}^{(j)})$
            \LComment{Evaluate properties using trained regressors}
            \State $\mathbf{s}^{(j)} \gets \mathbf{f}(\tilde{x}^{(j)})$
            \State Update $\mathcal{P}^\star$ with $\tilde{\mathbf{x}}^{(j)}$ if non-dominated
            \LComment{Compute reward based on Pareto dominance}
            \State $\mathbf{r}^{(j)} \gets \textsc{ComputeReward}(\mathbf{s}^{(j)}, \mathcal{P}^\star)$
        \EndFor
        
        \State \colorbox{gray!20}{\textbf{Backpropagation:}}
        \LComment{Update ancestor node statistics with child rewards}
        \State $\textsc{Backpropagate}(\mathbf{x}_{\text{leaf}}, \{\mathbf{r}^{(j)}\})$
    \EndFor
    
    \State \textbf{return} $\mathcal{P}^\star$
\end{algorithmic}
\end{algorithm}
\end{document}